\documentclass[12pt,a4paper]{article}
\usepackage[german, english]{babel}
\usepackage{a4wide}
\usepackage{amsmath}
\usepackage{amssymb}
\usepackage{ifthen}
\usepackage{epsfig}
\begin{document}
\newcounter{fig}   
\newcommand{\lbfig}[1]{\refstepcounter{fig}\label{#1}} 
\newcommand{\vacuum}{|0 \rangle}
\newcommand{\hmu}{\hat\mu}
\newcommand{\vacuumt}{\langle 0|}
\newcommand{\flavor}{|\text{flavor}\rangle} 
\newcommand{\diag}{\operatorname{diag}}
\newcommand{\ftb}{\bar{g}} \newcommand{\ft}{g}
\newcommand{\Qb}{\bar{Q}} \newcommand{\Q}{Q} \newcommand{\fb}{\bar{f}}
\newcommand{\f}{f} \newcommand{\qb}{\bar{q}} \newcommand{\q}{q}
\newcommand{\pb}{\bar{p}} \newcommand{\p}{p}
\newcommand{\psib}{\bar{\psi}} \newcommand{\phib}{\bar{\phi}}
\newcommand{\vb}{\bar{v}} \newcommand{\U}{U} \newcommand{\Ub}{\bar{U}}
\newcommand{\V}{V} \newcommand{\Vb}{\bar{V}} \newcommand{\J}{J}
\newcommand{\Jb}{\bar{J}} \newcommand{\dirac}{\not \hspace{-1.2mm} D}
\renewcommand{\det}{\operatorname{Det}}
\newcommand{\tr}{\operatorname{Tr}}
\newcommand{\sgn}{\operatorname{sgn}}
\newcommand{\vol}{\operatorname{Vol}}
\newcommand{\const}{\text{const.}}
\newcommand{\length}{\text{[length]}} \newcommand{\n}{\mathcal{N}}
\newcommand{\arcosh}{\operatorname{arcosh}}
\newcommand{\Gr}[2][]{\ensuremath{ \mathrm{#2}
\ifthenelse{\equal{#1}{}} {} {(#1)} } } \newcommand{\lie}{\mathfrak}
\newcommand{\gr}[2][]{\ensuremath{ \lie{#2} \ifthenelse{\equal{#1}{}}
{} {(#1)} } }     
\newcommand{\minuspt}{\\[-4pt]}
\newcommand{\abb}[3]{\ensuremath{ \ifthenelse{\equal{#1}{}}{}{#1:} #2
\mapsto #3}} 
\newcommand{\abbildung}[6][]{\begin{align}
\ifthenelse{\equal{#1}{}}{}{\label{#1}}
\ifthenelse{\equal{#2}{}}{}{#2:} #3 & \rightarrow   #4  \minuspt #5 &
\mapsto #6  \notag \end{align}} 
\def\ad{\mathop{\rm ad}\nolimits}
\def\Ad{\mathop{\rm Ad}\nolimits}
\def\Det{\mathop{\rm Det}\nolimits}
\def\un{\mathop{\mathfrak{u}(N)}\nolimits}
\def\eps{\epsilon}
\def\vareps{\varepsilon}
\def\Ker{\mathop{\rm Ker}\nolimits}
\def\tr{{\rm tr}}
\def\Tr{{\rm Tr}}
\def\re{{\rm Re\;}}
\def\im{{\rm Im\;}}
\def\mi{{\rm i}}
\def\e{\mathop{\rm e}\nolimits}
\def\sq3{\sqrt{3}}
\def\isq3{\frac{1}{\sqrt{3}}}
\def\sqn{\sqrt{N}}
\def\defi{\stackrel{\rm def}{=}}
\def\t2{{\Bbb T}^2}
\newcommand{\ahalf}{\frac{a}{2}}
\newcommand{\treehalf}{\frac{3a}{2}}
\newcommand{\Ss}{\scriptstyle}
\newcommand{\mG}{\mathsf{G}}
\newcommand{\dr}{\ensuremath{\mathsf{R}}}
\newcommand{\di}{\ensuremath{\mathsf{I}}}
\newcommand{\npmuh}{\ensuremath{({\scriptstyle n+\frac{\hat\mu}{2}})}}
\newcommand{\nppmuh}{\ensuremath{({\scriptstyle n'+\frac{\hat\mu}{2}})}}
\newcommand{\nmmuh}{\ensuremath{({\scriptstyle n-\frac{\hat\mu}{2}})}}
\newcommand{\npnuh}{\ensuremath{({\scriptstyle n+\frac{\hat\nu}{2}})}}
\newcommand{\nmnuh}{\ensuremath{({\scriptstyle n-\frac{\hat\nu}{2}})}}
\newcommand{\npmu}{\ensuremath{( n+\hat\mu)}}

\begin{titlepage}

  \title{$(1+1)$--dimensional Baryons from the ${\rm
      SU}(N)$~Color--Flavor Transformation}
\author{
J.~Budczies$^a$, S.~Nonnenmacher$^b$, Ya.~Shnir$^a$, 
 and M.R.~Zirnbauer$^a$ \\ [2mm]
{\it {\footnotesize 
$^a$Institut f\"ur Theoretische Physik, Universit\"at zu K\"oln, 
50937 K\"oln, Germany}}\\
{\it {\footnotesize$^b$Service de Physique Th\'eorique,  CEA-Saclay, 
91191 Gif-sur-Yvette Cedex, France
}}
\date{~}
}
\maketitle 
\begin{abstract}
  The color--flavor transformation, an identity that connects two
  integrals, each of which is over one of a dual pair of Lie groups
  acting in the fermionic Fock space, is extended to the case of the
  special unitary group.  Using this extension, a toy model of lattice
  QCD is studied: $N_f$ species of spinless fermions interacting with
  strongly coupled ${\rm SU}(N_c)$ lattice gauge fields in $1+1$
  dimensions.  The color--flavor transformed theory is expressed in
  terms of gauge singlets, the meson fields, organized into sectors
  distinguished by the distribution of baryonic flux.  A comprehensive
  analytical and numerical search is made for saddle--point
  configurations of the meson fields, with various topological
  charges, in the vacuum and single--baryon sectors.  Two definitions
  of the static baryon on the square lattice, straight and zigzag, are
  investigated.  The masses of the baryonic states are estimated using
  the saddle--point approximation for large $N_c$.
\end{abstract}
\end{titlepage}

\section{Introduction}

In quantum chromodynamics (QCD), a hierarchy of scales is provided by
$\Lambda_{\chi} \sim ~ 1 ~{\rm GeV}$, the scale of chiral symmetry
breaking, and $\Lambda_{\rm QCD}\sim ~0.18 ~ {\rm GeV}$, defined as
the location of the Landau pole of the one--loop beta function.  The
running coupling constant increases from weak to strong coupling as
the momentum scale is lowered from the perturbative regime above
$\Lambda_{\chi}$ down to $\Lambda_{\rm QCD}$.

In the past two decades a great deal was learned about the
non--perturbative structure of QCD at scales between $\Lambda_{\chi}$
and $\Lambda_{\rm QCD}$.  The guiding idea was to construct
low--energy effective theories which encode the symmetries of the
fundamental QCD Lagrangian.  To obtain these effective theories, one
may start from full QCD, and integrate out the high--energy degrees of
freedom (quarks and gluons) in order to produce a low--energy
effective action in terms of mesons and baryons.  In this way it was
possible to recover the chiral Lagrangian \cite{Meis,Schur,Diak,DP}
that had been introduced phenomenologically by Weinberg \cite{Wein}.

In a more recent development, it was shown \cite{Rebbi} how to extract
the effective long--distance degrees of freedom by starting from the
lattice \cite{Wilson74,Creutz} formulation of QCD.  In that approach
it is assumed that the long--distance physics of lattice QCD (LQCD)
can be described by a strongly coupled lattice theory.  From the
latter, one gets the continuum chiral Lagrangian by expanding the
effective action in powers of the lattice spacing and external
momenta.  All the terms of the Gasser--Leutwyler continuum effective
Lagrangian \cite{gl} can be recovered in this way \cite{Levi}.  The
lattice formulation is, however, deficient in one respect: by the
technical difficulties with chiral symmetry for lattice fermions, the
chiral anomaly is lost, i.e.~for $N_f$ massless quark flavors the
chiral symmetry of the lattice effective theory is ${\rm U}(N_f)$
rather than ${\rm SU}(N_f)$.

This type of approach was initiated in \cite{Kluberg,Smith}; it relied
on a ``bosonization'' of the strong--coupling LQCD action, and a
large--$N_c$ or large--dimension expansion.  Technically, the heart of
the method is the computation of integrals over the group
$\Gr[N_c]{SU}$ with Haar measure, weighted by $\e^{-S(U)}$.  Some
general results for such integrals have recently been reviewed in
\cite{Balantekin}.

A few years ago, an alternative kind of bosonization scheme was
introduced \cite{circular}, relying on a mathematical formalism later
called the ``color--flavor transformation'' \cite{icmp97}.  This
transformation relates two different formulations of a certain class
of theories.  In condensed matter theory, the transformation has found
a number of applications, among others to the random flux model
\cite{as}.

The color--flavor transformation in its original version applies to
the gauge group ${\rm U}(N_c)$.  For this group, all gauge singlets
are of ``mesonic'' (or quark--antiquark) type.  In order for baryons
to appear, one needs to replace ${\rm U}(N_c)$ by the special unitary
group ${\rm SU}(N_c)$.  In Section \ref{sec:cft} of the present paper
we extend the color--flavor transformation to ${\rm SU}(N_c)$, by
decomposing the (colorless) flavor sector of Fock space into
disconnected subsectors labeled by the baryonic charge.

In Sections \ref{vac-config}--\ref{sec:zigzag} we apply the formalism
to a toy model of LQCD: $N_f$ species of spinless fermions interacting
with strongly coupled ${\rm SU}(N_c)$ lattice gauge fields in $1+1$
dimensions.  The color--flavor transformation yields a dual
representation of this non--Abelian model.  Combining numerical
computations with analytical considerations, we conduct a
comprehensive search for saddle--point configurations in various
baryonic sectors with different topological properties.  We use these
configurations (without fluctuation corrections) to estimate the mass
of a single baryon in our model.  In doing so we ignore the
Mermin--Wagner--Coleman theorem (asserting that spontaneous breaking
of continuous global symmetries does not occur in $1+1$ dimensions),
by assuming the pattern of chiral symmetry breaking that is known to
occur in the physical case of $3+1$ dimensions.

After the present work had been completed, we learned that the 
${\rm SU}(N)$ generalization of the color--flavor transformation 
has also been worked out by Schlittgen and Wettig \cite{su-rivals}.

\section{Color--Flavor Transformation for ${\rm SU}(N_c)$}\label{sec:cft}

\subsection{Group action on fermionic Fock space}\label{definitions}

In this section we set up some algebraic structures, which are needed
to establish the ``color--flavor'' transformation for the special
unitary group.  Our discussion follows the line of reasoning of
Ref.~\cite{circular} but is somewhat simpler, as we do not need the
superalgebraic framework employed there.

We start by considering a set of fermionic creation and annihilation
operators $\bar f_A^i$ and $f_A^i$, which obey the canonical
anticommutation relations
\begin{displaymath}
  \{ \f^i_A, \f^j_B \} = 0 \;,\qquad \{ \fb^i_A, \fb^j_B \} = 0 \;,
  \qquad \{ \f^i_A, \fb^j_B \} = \delta_{AB} \delta^{ij} \;.
\end{displaymath}
The lower index takes the values $+a$ or $-a$, with range $a = 1,
\ldots, N_f$, and the upper index takes the values $i = 1, \ldots,
N_c$.  Having QCD in mind, we interpret the operators $\bar f_{+a}^i$
and $\bar f_{-a}^i$ as creation operators for ``quarks'' and
``antiquarks'' respectively; the index $i$ corresponds to the gauge
(or color) degrees of freedom and the index $a$ labels the different
quark flavors.  (The quarks are regarded here as being spinless.)  The
operators $f_A^i$ and $\bar f_A^i$ act on a Fock space with vacuum
$\vacuum$ and its conjugate $\vacuumt$, by $\f^i_A \vacuum = 0$ and
$\vacuumt \fb^i_A =0 $ for all $A$ and $i$.

We next consider the set of quadratic operators $E_{AB}^{ij}$ defined
by
\begin{alignat}{2}
  E_{+a,+b}^{ij} &= \fb_{+a}^i \f_{+b}^j \;, \qquad &E_{+a,-b}^{ij} &=
  \fb_{+a}^i \fb_{-b}^j \;, \nonumber \\ E_{-a,+b}^{ij} &= \f_{-a}^i
  \f_{+b}^j \;, &E_{-a,-b}^{ij} &= \f_{-a}^i \fb_{-b}^j \;. \nonumber
\end{alignat}
The ${\mathbb C}$--linear span of these operators has the structure of
a complex Lie algebra, $\mathfrak{G}$.  More precisely, the operators
$E_{AB}^{ij}$ obey the commutation relations of a set of canonical
generators of the Lie algebra $\gr[2N_f N_c]{gl}$:
\begin{displaymath}
  [E_{AB}^{ij}, E_{CD}^{kl}] = \delta^{jk} \delta_{BC} E^{il}_{AD} -
  \delta^{li} \delta_{DA} E_{CB}^{kj} \;.
\end{displaymath}
Thus we have a Lie algebra isomorphism from $\gr[2N_fN_c]{gl}$
(i.e.~the space of complex matrices of size $2N_f N_c \times 2N_f
N_c$, with the Lie bracket given by the commutator) to $\mathfrak{G}$:
\begin{eqnarray*}
  t : \gr[2N_fN_c]{gl} &\to& \mathfrak{G} \;, \\ m &\mapsto& t_m :=
  \sum_{ij,AB} m_{AB}^{ij} E_{AB}^{ij} \;.
\end{eqnarray*}
This isomorphism lifts to an isomorphism of the corresponding complex
groups:
\begin{eqnarray}
  T : {\rm GL}(2N_f N_c) &\to& G \;, \nonumber \\ M = \exp(m)
  &\mapsto& T_M = \exp(t_m) \;, \label{liegroup}
\end{eqnarray}
which forms a (reducible) representation of ${\rm GL}(2N_f N_c)$ on
Fock space.  The representation is single--valued (which means there
are no ${\rm U}(1)$ obstructions from the multi-valuedness of the
logarithm) as the spectrum of each operator $E_{AA}^{ii}$ is the set
$\{ 0, 1 \}$.

The Lie algebra $\gr[2N_f N_c]{gl}$ has two subalgebras $\gr[N_c]{gl}$
and $\gr[2N_f]{gl}$ which are embedded in a natural way: a matrix $X
\in \gr[N_c]{gl}$ is identified with $I_{2N_f}\otimes X$, and a matrix
$Y \in \gr[2N_f]{gl}$ with $Y \otimes I_{N_c}$.  Through these
embeddings, $\gr[N_c]{gl}$ and $\gr[2N_f]{gl}$ form a pair of maximal
commuting subalgebras of $\gr[2N_f N_c]{gl}$, also known as a ``dual
pair'' \cite{Howe}.  The subgroups ${\rm GL}(N_c)$ and ${\rm GL}
(2N_f)$ are embedded into ${\rm GL}(2N_f N_c)$ in the same way.  Their
adjoint action on the fermionic creation and annihilation operators is
described in Appendix \ref{appA}.

We define the \emph{color group} to be the subgroup ${\rm SU}(N_c)$ of
${\rm GL}(N_c)$, and the \emph{flavor group} to be the subgroup ${\rm
  U}(2N_f)$ of ${\rm GL}(2N_f)$.  ${\rm GL}(N_c)$ contains an extra
${\rm U}(1)$ subgroup which lies outside the color group and, being
generated by the unit matrix, commutes with the whole group ${\rm
  GL}(2 N_f N_c)$.  This ${\rm U}(1)$ is generated by $\hat Q + N_f$
where
\begin{equation}\label{Casimir}
  \hat{Q} = \frac{1}{N_c} \sum_{A,i}E_{AA}^{ii} - N_f = \frac{1}{N_c}
  \sum_{a,i}(\fb_{+a}^i \f_{+a}^i - \fb_{-a}^i \f_{-a}^i)
\end{equation}
counts the difference between the number of particles and
antiparticles: $\hat Q = \frac{1}{N_c}(N_+ - N_-)$.  In contrast, the
operator giving the total number of particles, $\hat{N} = \sum_{a,i}
(\fb_{+a}^i \f_{+a}^i + \fb_{-a}^i \f_{-a}^i)$, does not commute with
the generators of $\gr[N_f]{gl}$.  We will call $\hat Q$ the
\emph{baryon charge operator}.

\subsection{From color group integrals to flavor group integrals}

Let $\psi_A^i$ and $\psib_A^i$ be two independent sets of Grassmann
variables, referred to as ``quark fields'', and consider the color
group integral
\begin{equation}\label{genfunction}
  \mathcal{Z}(\psi, \psib) = \int_{\Gr[N_c]{SU}} dU \exp(\psib_{+a}^i
  U^{ij} \psi_{+a}^j + \psib_{-b}^i \bar{U}^{ij} \psi_{-b}^j) \;.
\end{equation}
The Haar measure $dU$ of $\Gr[N_c]{SU}$ is understood to be normalized
by $\int_{{\rm SU}(N_c)} dU = 1$.  We also adopt the convention that
repeated occurrence of an index implies summation.

The color--flavor transformation will replace the integral
(\ref{genfunction}) by an integral over the flavor group \Gr[2N_f]{U}.
A key step in doing the transformation is to interpret $\mathcal{Z}
(\psi, \psib)$ as the matrix element of an operator $\mathcal{P}$ that
projects on the \emph{colorless sector} (or flavor sector) of Fock
space.  This sector is the subspace of all states $\flavor$ which are
invariant under the color group: $T_U \flavor = \flavor$ for all $U
\in \Gr[N_c]{SU}$.

The first step towards the color--flavor transformation is to express
the projector $\mathcal{P}$ as
\begin{equation}
  \mathcal{P} =  \int_{\Gr[N_c]{SU}} dU \, T_U \;.
\end{equation}
Let us now introduce the fermion coherent states
\begin{equation}
  \langle \bar{\Psi} | = \vacuumt \exp(\psib_{-a}^i \f_{-a}^i +
  \psib_{+a}^i \f_{+a}^i) \;, \qquad | \Psi \rangle = \exp(\fb_{-a}^i
  \psi_{-a}^i + \fb_{+a}^i \psi_{+a}^i) \vacuum \;.
\end{equation}
By making use of the first set of relations in Appendix A, it is
straightforward to show that
\begin{displaymath}
  \langle \bar{\Psi} |T_U| \Psi \rangle = \exp(\psib_{+a}^i U^{ij}
  \psi_{+a}^j + \psib_{-b}^i \bar{U}^{ij} \psi_{-b}^j)
\end{displaymath}
for $U \in {\rm SU}(N_c)$.  This yields the simple formula
\begin{equation}\label{Zformula}
  \mathcal{Z}(\psi, \psib) = \langle \bar{\Psi} | \mathcal{P} | \Psi
  \rangle \;.
\end{equation} 

To express $\mathcal{Z}(\psi, \bar{\psi})$ as an integral over the
flavor group, we will derive an alternative representation of the
projector $\mathcal{P}$, as an integral over coherent states of the
flavor sector.

\subsection{The flavor sector}

The subspace of states in Fock space which are invariant under
$\Gr[N_c]{U}$ was described in \cite{circular}.  It consists of the
vacuum and of \emph{mesonic} excitations on top of it.  The prototype
of such an excitation is the ``one--meson'' state
\begin{displaymath}
  | m_{ab} \rangle = \sum_i E^{ii}_{+a,-b} \vacuum = \sum_i \fb^i_{+a}
  \fb^i_{-b} \vacuum \;.
\end{displaymath}
By the multiple action of the $\gr[2N_f]{gl}$ generators $E^{ii}_{
  +a,-b}$ (where we have gone back to using the summation convention),
one can build states containing up to $N_c N_f$ mesons, with different
flavors.  These states are automatically $\Gr[N_c]{U}$--invariant;
conversely, all $\Gr[N_c]{U}$--invariant states are linear
combinations of such multi--meson states.  The group $\Gr[2N_f]{U}$
acts irreducibly on this invariant subspace.

The set of ${\rm SU}(N_c)$--invariant states is larger.  To obtain it,
one relaxes the constraint $\hat Q|\psi\rangle = 0$.  Thus there exist
colorless sectors of Fock space on which the central generator $\hat
Q$ takes a non--zero value.  These sectors contain the \emph{baryons},
which are totally antisymmetric combinations of $N_c$ quarks.  A
baryon with flavors $a_1, \ldots, a_{N_c}$ is defined as
\begin{equation}\label{1baryon}
  | b_{A_1 \ldots A_{N_c}} \rangle = \frac{1}{N_c!} \varepsilon_{i_1
    \ldots i_{N_c}} \fb_{A_1}^{i_1} \cdot \ldots \cdot \fb_{A_{N_c}}^
  {i_{N_c}} | 0 \rangle \;,
\end{equation}
where the $A_k = \pm a_k$ are taken either all positive (baryon), or
all negative (antibaryon).  A matrix $g \in \Gr[N_c]{GL}$ acts on this
state simply by multiplication with ${\rm Det}(g)$ (resp.~${\rm Det}^{
  N_f-1} (g)$.  Therefore, the state is invariant under the color
group $\Gr[N_c]{SU}$.

The above baryon (resp.~antibaryon) is an eigenstate of the baryon
charge operator $\hat Q$ with eigenvalue $+1$ (resp.~$-1$).  Acting on
it with the generators $E^{ii}_{AB}$ of the flavor algebra $\gr[2N_f]
{gl}$, one builds other colorless states with the same baryon number,
which form an irreducible subspace for $\Gr[2N_f]{U}$: the one--baryon
(resp.~one--antibaryon) sector.

The one--baryon sector can be generated from the state (\ref{1baryon})
with all $a_j = 1$.  One can similarly build $Q$--baryon
(resp.~$Q$--antibaryon) states from
\begin{equation}  \label{startstates}
  | B_Q \rangle = \prod_{a=1}^Q \fb_{+a}^1 \cdot ... \cdot \fb_{+a}^
  {N_c} \vacuum \;, \quad | B_0 \rangle = \vacuum \;, \quad | B_{-Q}
  \rangle = \prod_{a=1}^Q \fb_{-a}^1 \cdot ... \cdot \fb_{-a}^{N_c}
  \vacuum \;.
\end{equation}
The values of the baryon charge range from $-N_f$ to $N_f$, according
to Pauli's exclusion principle.  As with $Q = \pm 1$, acting on $| B_Q
\rangle$ with the algebra $\gr[2N_f]{gl}$ builds the full $Q$--baryon
part of the flavor sector, so the group $\Gr[2N_f]{U}$ acts
irreducibly on this part.  This can be proved by using the dual--pair
property of the subalgebras $\gr[2N_f]{gl}$ and $\gr[N_c]{gl}$, as
exposed in \cite{Howe}.

To summarize, the flavor sector of Fock space decomposes into $2N_f +
1$ subsectors, characterized by their baryon charges $Q$.  Each sector
carries an irreducible unitary representation of the flavor group
$\Gr[2N_f]{U}$.

\subsection{Coherent states}

Having decomposed the flavor sector as described above, we can now
express the projector ${\cal P}$ in a different way.  For this purpose
we will use coherent states, in the spirit of Perelomov
\cite{Perelomov}.  On each subsector with a fixed baryon charge $Q$,
we consider the \emph{generalized coherent states} built by the action
of $G \equiv \Gr[2N_f]{U}$ on the reference state $|B_Q\rangle$,
i.e.~the states
\begin{equation}
  \quad\forall g\in G,\ \forall Q = -N_f, \ldots, N_f : \quad | g_Q
  \rangle \defi T_g|B_Q\rangle,\quad \langle g_Q\mid\defi \langle B_Q
  |T_g^\dagger \;.
\end{equation}
The crucial property of coherent states we will now use, is that they
supply a resolution of unity.  Because of the irreducibility of the
$\Gr[2N_f]{U}$ action on each $Q$--subsector, the operator
\begin{equation}
  \mathcal{P}_Q \defi \alpha_Q \int_{G} dg \, |g_Q\rangle\langle g_Q |
\end{equation}
coincides with the orthogonal projector on that subsector, the only
provision being that the normalization constant $\alpha_Q$ be chosen
appropriately.  Indeed, the operator $\mathcal{P}_Q$ trivially
commutes with every element of the flavor group; Schur's lemma then
ensures that it is proportional to the identity on each irreducible
space of this group, therefore on each subsector with fixed baryonic
charge.  Owing to orthogonality, ${\cal P}_Q$ vanishes on all
subsectors with $Q'\neq Q$, whereas it is the identity on the
$Q$--subsector if we take
\begin{equation}
  \alpha_Q = \left( \int_{G} dg \, |\langle B_Q | T_g | B_Q \rangle|
    ^2 \right)^{-1} \;.
\label{alphaQ}
\end{equation}
Some particular values of the constant (namely $\alpha_0$, $\alpha_{
  \pm 1}$) are computed in Appendix \ref{normalisation}.  

For the matrix element (\ref{Zformula}) of the projector $\mathcal{P}$
on the full flavor sector,
\begin{displaymath}
  \mathcal{P} = \bigoplus_{Q=-N_f}^{N_f} \mathcal{P}_Q \;,
\end{displaymath}
we now have a new representation:
\begin{equation}\label{genfunc2}
  \mathcal{Z}(\psi,\psib) = \sum_{Q=-N_f}^{N_f}\alpha_Q \int_{G} dg
  \,\langle \bar{\Psi}|g_Q\rangle \langle g_Q |\Psi \rangle \;.
\end{equation}
To compute the overlaps $\langle \bar\Psi | g_Q \rangle$ and $\langle
g_Q | \Psi \rangle$, it is convenient to use a Gauss decomposition of
$G = {\rm U}(2N_f)$: almost any matrix $g = \begin{pmatrix} A&B\\ C&D
\end{pmatrix} \in G$ can be factored as
\begin{equation}
  \begin{pmatrix} A & B \\ C & D \end{pmatrix} = 
  \begin{pmatrix} 1 & Z \\ 0 & 1 \end{pmatrix}
  \begin{pmatrix} \tilde{A} & 0 \\ 0 & D \end{pmatrix}
  \begin{pmatrix} 1 & 0 \\ \tilde{Z} & 1 \end{pmatrix},
\label{decomposition}
\end{equation}
where the relations $Z = BD^{-1}$, $\tilde Z = D^{-1}C$, and $\tilde A
= A - BD^{-1}C$ hold.  The decomposition becomes singular if $D$ does,
but this happens only on a submanifold of codimension one (and hence
measure zero) of $G$.  The unitarity of $g$ implies $\tilde Z = -
D^\dagger Z^\dagger {A^\dagger}^{-1}$ and allows to write the central
matrix in the form
\begin{equation}
\label{ADdecompo}
  \begin{pmatrix} \tilde{A} & 0 \\ 0 & D \end{pmatrix} = 
  \begin{pmatrix} (1+ Z Z^\dagger)^{1/2} & 0 \\
                   0 & (1+ Z^\dagger Z)^{-1/2} \end{pmatrix} 
  \begin{pmatrix} \mathcal{U} & 0 \\ 0 & \mathcal{V} \end{pmatrix} \;.
\end{equation}
$\mathcal{U}$ and $\mathcal{V}$ are unitary, so $\begin{pmatrix}
  \mathcal{U} & 0 \\ 0 & \mathcal{V} \end{pmatrix}$ is an element of
the diagonal $\Gr[N_f]{U}\times\Gr[N_f]{U}$ subgroup of $G$, which we
call $H$.  It can thus be shown that the elements $g$ of an open dense
subset of $G$ are in one-to-one correspondence with the triplets $(Z,
\cal{U},\cal{V})$, where the pair $\diag(\cal{U}, \cal{V})$ is an
element of $H$, while $Z$ represents a point in the coset space $G/H$
and can be any complex $N_f\times N_f$ matrix.  Moreover, the Haar
measure $dg$ of $G$ factorizes as
\begin{equation}\label{groupintegral}
  \int_G dg = \int_{G/H} d(gH) \int_H dh = \int_{\mathbb{C}^{N_f
      \times N_f}} d\mu(Z,Z^\dagger)\int_{H}d{\cal U}\,d{\cal V} \;.
\end{equation}
Both $d\cal{U}$ and $d\cal{V}$ are normalized Haar measures on
$\Gr[N_f]{U}$, and
\begin{displaymath}
  d\mu(Z, Z^\dagger) = C_{N_f} \det(1 + Z Z^\dagger)^{-2N_f} \,
  \prod_{i,j} dZ_{ij} d\bar Z_{ij}
\end{displaymath}
is the normalized invariant measure on $G/H$.  The normalization
factor $C_{N_f}$ is computed in Appendix \ref{normalisation}; see
Eq.~(\ref{CNf}).

We now explain how to use this decomposition to compute the overlaps.
The Gauss decomposition (\ref{decomposition}) carries over to any
representation of $G$, so for every $g\in G$ we can write the operator
$T_g$ as
\begin{equation} 
  T_g = T_\zeta \, T_{\diag(\tilde{A},D)} \, T_{\tilde{\zeta}} \;,
\end{equation} 
where $\zeta = \begin{pmatrix} 1 & Z \\ 0 & 1 \end{pmatrix}$ and
$\tilde{\zeta} = \begin{pmatrix} 1 & 0 \\ \tilde{Z} & 1
\end{pmatrix}$.  According to the relations given in Appendix
\ref{appA}, the factors $T_\zeta$ and $T_{\tilde{\zeta}}$ act
trivially on the reference states:
\begin{equation}
  \forall Q = - N_f, \ldots, N_f : \qquad T_{\tilde{\zeta}} | B_Q
  \rangle = | B_Q \rangle, \quad \langle B_Q | T_{\zeta} = \langle B_Q
  | \;.
\end{equation}
The action of the block-diagonal operator is slightly more subtle.
Using the third set of relations given in Appendix A, we get
\begin{align}
  T_{\diag(\tilde{A}, D)} \vacuum &= (\det D)^{N_c}\vacuum \;,
  \nonumber \\ T_{\diag(\tilde{A}, D)}|B_1\rangle &= (\det
  D)^{N_c}\prod_{i=1}^{N_c} \tilde{A}_{a 1}\fb_{+a}^{i}\vacuum \;,
  \nonumber \\ T_{\diag(\tilde{A}, D)}|B_{-1}\rangle &= (\det
  D)^{N_c}\prod_{i=1}^{N_c} (D^{-1})_{1 a}\fb_{-a}^{i}\vacuum \;.
  \nonumber
\end{align}
(To make sense of these formulas one must remember that we are using
the summation convention: the flavor index $a$ under the product is
understood to be summed over.) These formulas directly yield the
desired overlaps with $\langle {\bar \Psi}|$:
\begin{align}
  \langle {\bar \Psi}| g_0\rangle &= (\det D )^{N_c} \prod_{i=1}^{N_c}
  \exp( \psib^i_{+a} Z_{ab} \psib^i_{-b}) \;, \nonumber \\ \langle
  {\bar \Psi} | g_1 \rangle &= (\det D )^{N_c} \prod_{i=1}^{N_c}
  \psib^i_{+c} \tilde{A}_{c1} \exp( \psib^i_{+a} Z_{ab} \psib^i_{-b})
  \;, \nonumber \\ \langle {\bar \Psi} | g_{-1} \rangle &= (\det D
  )^{N_c} \prod\limits_{i = 1}^{N_c} D^{-1}_{1c} \psib^i_{-c} \exp(
  \psib^i_{+a} Z_{ab} \psib^i_{-b}) \;, \nonumber
\end{align}
as well as the overlaps with $|\Psi \rangle$:
\begin{align}
  \langle g_0 | \Psi \rangle &= (\det D^\dagger )^{N_c}
  \prod_{i=1}^{N_c} \exp(\psi^i_{-a} {Z^\dagger}_{ab} \psi^i_{+b}) \;,
  \nonumber \\ \langle g_1 | \Psi \rangle &= (\det D^\dagger )^{N_c}
  \prod_{i=1}^{N_c} {\tilde{A}^\dagger}_{1c} \psi^i_{+c}
  \exp(\psi^i_{-a} {Z^\dagger}_{ab} \psi^i_{+b}) \;, \nonumber \\ 
  \langle g_{-1} | \Psi \rangle &= (\det D^\dagger)^{N_c}
  \prod\limits_{i=1}^{N_c} \psi^i_{-c} {(D^{-1})}^\dagger_{c1}
  \exp(\psi^i_{-a} {Z^\dagger}_{ab} \psi^i_{+b}) \;. \nonumber
\end{align}
The overlaps with the coherent states $| g_Q \rangle$ containing more
than one baryon ($|Q| > 1$) can be computed in the same way; in front
of the exponential factors, there will be $|Q|$ similar products, with
flavor indices $1, \ldots, |Q|$.

We now insert the above expressions for the overlaps into
(\ref{genfunc2}), and use the factorization (\ref{groupintegral}) to
arrive at an integral over triples ($Z, {\cal U}, {\cal V}$).  Leaving
the $Z$--integral for later, we next carry out the integrations over
the unitary matrices ${\cal U}$ and ${\cal V}$.  They enter in the
overlaps via the matrix elements of $\tilde A$ and $D$; see
Eq.~(\ref{ADdecompo}).  To simplify the notation, we first perform a
flavor rotation on the Grassmann fields:
\begin{alignat}{2}
  \phi_{+b}^i &= (\sqrt{1+ZZ^\dagger})_{ba} \psi_{+a}^i \;, \quad
  \phi_{-b}^i &= \psi_{-a}^i (\sqrt{1+ Z^\dagger Z})_{ab} \;,
  \nonumber \\ \phib_{+b}^i &= \psib_{+a}^i (\sqrt{1+Z
    Z^\dagger})_{ab} \;, \quad \phib_{-b}^i &= (\sqrt{1+Z^\dagger
    Z})_{ba} \psib_{-a}^i \;. \nonumber
\end{alignat}
The integrals we need to compute then read as follows (assuming $Q >
0$):
\begin{align}
  \chi_Q(\phib_+, \phi_+) &\defi \alpha_Q \int\limits_{{\rm U}(N_f)}\!
  d \mathcal{U} \prod_{c=1}^Q \prod_{i=1}^{N_c} (\phib^i_{+a}
  \mathcal{U}_{ac}) (\phi_{+b}^i \mathcal{U} ^{-1}_{cb}) \;, \\ 
  \chi_{-Q}(\phib_-, \phi_-) &\defi \alpha_{-Q} \int\limits_{{\rm
      U}(N_f)}\!  d \mathcal{V} \prod_{c = 1}^Q\prod_{i=1}^{N_c}
  (\phib^i_{-a} \mathcal{V}^{-1}_{ca}) (\phi_{-b}^i \mathcal{V}_{bc})
  \;.
\end{align}
We also set $\chi_0 \equiv \alpha_0$, and $\chi_{Q} (\bar\psi,\psi;Z)
\equiv \chi_{Q} (\bar\phi_+,\phi_+)$, and $\chi_{-Q} (\bar\psi,\psi;Z)
\equiv \chi_{-Q}(\bar\phi_-,\phi_-)$.  The function $\chi_1(\bar\phi_+
,\phi_+)$ will play a distinguished role in the lattice gauge theory
application in Section \ref{latticegauge}, and we therefore evaluate
it explicitly in the next subsection.

The integrations over $H$ having been done, we are left with an
integral over $G/H$, i.e.~over a $Z$--dependent integrand, in each
$Q$--subsector.  Putting everything together, we finally arrive at the
following identity:
\begin{eqnarray}
  &&\int_{\Gr[N_c]{SU}} dU \exp(\psib_{+a}^i U^{ij} \psi_{+a}^j +
  \psib_{-b}^i \bar{U}^{ij} \psi_{-b}^j) \nonumber \\ &=& \sum_{Q = -
    N_f}^{N_f}\ \int_{\mathbb{C}^{N_f \times N_f}}\! d\mu(Z,
  Z^\dagger)\; \chi_Q(\bar\psi,\psi;Z) \, \frac{\exp(\psib_{+a}^i
    Z_{ab} \psib_{-b}^i + \psi_{-b}^j {Z^\dagger}_{ba} \psi_{+a}^j)}
  {\det(1+Z Z^\dagger)^{N_c}} \;, \label{cft}
\end{eqnarray}
which is called the color--flavor transformation for ${\rm SU}(N_c)$,
and is the central result of the present section.  Note that the
right-hand side of the transformation has the attractive feature of
organizing the contributions according to the different baryonic
sectors.

An effective action in the bosonic variable $Z$ can be obtained by
doing the (Gaussian) integral over the Grassmann fields.  This will be
done in a lattice gauge context in Section \ref{latticegauge}.

\subsection{Evaluation of $\chi_1$}\label{integralU}

In this subsection we evaluate the coefficient
\begin{displaymath}
  \chi_1(\phib_+, \phi_+) = \alpha_1 \int_{\Gr[N_f]{U}}
  d\mathcal{U}\ \prod_{i=1}^{N_c} (\phib^i_{+a} \mathcal{U}_{a1})
  (\phi^i_{+b} \bar{\mathcal{U}}_{b1}) \;.
\end{displaymath}
Only the first column of the unitary matrix $\mathcal{U}$ occurs in
the integrand, so the integral is effectively over a unit sphere in
$N_f$--dimensional complex space, ${\rm S}^{2N_f - 1} = {\mathbb C}
^{N_f} / {\mathbb R}_+$.  Parametrizing the latter by a complex vector
$z = (z_1, \ldots, z_{N_f})$ with unit norm $|z| = 1$, we have
\begin{displaymath}
  \chi_1(\phib_+, \phi_+) = \alpha_1 \frac{ \int_{|z| = 1}
    d\Omega(z,\bar{z}) \; \prod\limits_{i=1}^{N_c} (\phib^i_{+a} z_a)
    (\phi^i_{+b} \bar{z}_b)} {\int_{|z| = 1} d\Omega(z,\bar{z})} \;,
\end{displaymath}
where $d\Omega(z,\bar z)$ is a ${\rm U}(N_f)$--invariant measure on
the unit sphere $|z| = 1$.  By homogeneity in $z$ and $\bar z$, we may
use the trick of replacing the numerator and denominator by integrals
over $\mathbb{C}^{N_f}$, with a Gaussian weight function ${\rm
  e}^{-|z|^2}$ included in the integrands.  The answer then easily
follows from Wick's theorem:
\begin{eqnarray}
  \chi_1(\phib_+,\phi_+) &=& \alpha_1 \frac{(N_f-1)!} {(N_c+N_f-1)!}
  \sum_{\sigma \in \mathfrak{S}_{N_c}} \sgn \sigma \prod_{i=1}^{N_c}
  \phib^i_{+a} \phi_{+a}^{\sigma(i)} \nonumber \\ &=& \alpha_1
  \frac{(N_f-1)!}{(N_c+N_f-1)!} \sum_{\sigma \in \mathfrak{S}_{N_c}}
  \sgn \sigma \prod_{i=1}^{N_c} \psib_{+a}^i (1+ZZ^\dagger)_{ab}
  \psi_{+b}^{\sigma(i)} \;, \label{chi1}
\end{eqnarray}
where $\mathfrak{S}_{N_c}$ denotes the group of permutations of the
numbers $1, \ldots, N_c$.

\section{Color--flavor transformation on the lattice}
\label{latticegauge}

We consider a Euclidean $\Gr[N_c]{SU}$ gauge theory in $1+d$
dimensions placed on a hypercubic lattice with lattice constant $a$.
The fermions $\psi_b^i(n)$, with colors $i = 1, \ldots, N_c$ and
flavors $b = 1, \ldots, N_f$, are put on lattice sites labeled by $n =
(n_0, \ldots, n_d)$, while the gauge matrix variables $U \npmuh = \exp
\left( iag A_\mu(na + \frac{a\hat\mu}{2}) \right) \in {\rm SU}(N_c)$
are placed on lattice links $n + \hat\mu/2$ (we label links by their
\emph{middle points}), starting from sites $n$ in any of the
directions $\mu = 0, \ldots, d$.  In the limit of a strong gauge
coupling $g$, the gauge theory has the partition sum $\mathcal{Z} =
\int\prod\limits_n d\psi(n) d\psib(n) \mathcal{Z} (\psi,\psib)$ with
\cite{Kluberg}
\begin{equation}
  \mathcal{Z}(\psi,\psib) = \prod_{n} {\rm e}^{S_{m,\psi,\bar\psi}(n)}
  \prod_{\mu}\int_{\Gr[N_c]{SU}} d U\npmuh \ \e^{S_{U, \psi, \bar\psi}
    \npmuh}\;.
\end{equation}
The fermions on two neighboring sites $n$ and $n + \hat\mu$ are
coupled through the gauge fields on the connecting link $n + \hat\mu /
2$ in a gauge--invariant way:
\begin{equation}\label{basicaction}
  S_{U,\psi,\bar\psi} \npmuh = \frac{a^d}{2} \left( \psib^i_b(n)
    U^{ij}\npmuh \psi^j_b (n + \hat\mu) - \psib^j_b(n + \hat\mu)
    U^{\dagger ji}\npmuh \psi^i_b(n) \right) \;,
\end{equation}
while the (bare) quark mass $m$ couples the fermions diagonally
\begin{equation}
  S_{m,\psi,\bar\psi}(n) = a^{d+1} m\psib(n)\psi(n) \;.
\end{equation}
We are not going to worry here about the fermion doubling problem and
will restrict our considerations to this naive discretization of the
fermionic action.  Also, for simplicity we do not take into account
the spin degrees of freedom, leaving their inclusion for a future
publication.

We rescale the fermionic fields so as to absorb the prefactor $a^d/2$.
This just adds a global prefactor to $\mathcal{Z}$, and has no effect
on the physical quantities.  The ${\rm SU}(N_c)$--integral over $U$ on
each link is then identical to (\ref{genfunction}) after the following
substitutions:
\begin{displaymath}
  \psib_+ = \psib(n),\quad \psi_+ = \psi\npmu,\quad \psib_- =
  \psi(n),\quad \psi_- = \psib\npmu.
\end{displaymath}

On each link, we perform the color--flavor transformation (\ref{cft}),
thereby introducing a complex ``flavor matrix field'' $Z\npmuh$,
$Z^\dagger \npmuh$.  The outcome of the transformation reads
\begin{equation}
  \label{contributions}
  \mathcal{Z}(\psi,\psib) = \sum_{ \{ Q \} }
  \prod_{n}\e^{2am\bar\psi(n)\psi(n)} \prod_{\mu}
  \int_{\mathbb{C}^{N_f \times N_f}} d\mu(Z,Z^\dagger\npmuh) \,
  \chi^Q_{Z,\psi,\bar\psi}\npmuh \frac{{\rm e}^{ S_{Z,\psi,\bar\psi}
      \npmuh}} {\det(1+Z^\dagger Z\npmuh)^{N_c}}
\end{equation}
where the sum on the right--hand side extends over all possible
distributions $\{Q\}$ of baryonic charge (actually, baryonic flux)
over the links of the lattice.  The color--flavor transformed action
on a link $n + \hmu/2$ is
\begin{equation}
  S_{Z,\psi,\bar\psi} \npmuh = \psib_a^i(n) Z_{ab} \npmuh \psi_b^i(n)
  + \psib_b^j\npmu {Z^\dagger}_{ba} \npmuh \psi_a^j\npmu \;,
\end{equation}
and the $\chi$--coefficients are $\chi^0 \npmuh = \alpha_0$ and (with
$Q = Q\npmuh > 0$)
\begin{align}
  \chi^Q_{Z,\psi,\bar\psi}\npmuh &= \chi_{Q\npmuh} \left( \psib(n)
    \sqrt{1+ZZ^\dagger\npmuh}, \sqrt{1+ZZ^\dagger\npmuh} \psi \npmu
  \right) \;, \\ \chi^{-Q}_{Z,\psi,\bar\psi} \npmuh &= \chi_{-Q\npmuh}
  \left( \sqrt{1 + Z^\dagger Z\npmuh}\psi(n), \psib\npmu
    \sqrt{1+Z^\dagger Z\npmuh} \right) \;.
\end{align}
The fermions are now coupled through their \emph{flavor} indices,
whereas in the original action the coupling had been mediated by the
color degrees of freedom.  Moreover, the coupling has become
ultralocal: the fermions at a site $n$ couple only to one another, via
$Z\npmuh$, and so do the fermions at site $n+\hmu$, via $Z^\dagger
\npmuh$.  Correlations between neighbors are solely due to the
relation between $Z$ and $Z^\dagger$ by Hermitian conjugation.  A
graphical description of the change of coupling scheme is given in
Fig.~\ref{cf-1}.

The partition function (\ref{contributions}) is a sum over all
configurations of baryonic fluxes $\{Q\npmuh\}$.  For most of these
configurations, the Grassmann integral vanishes identically.  To see
that, we expand the integrand for a given configuration into a
polynomial in the Grassmann fields, and count (for each site $n$) the
number of fermions $\psi(n),\ \psib(n)$ in the various monomials:
\begin{itemize}
\item For every direction $\mu$, the coefficient $\chi^{Q}\npmuh$
  contains $N_c |Q|$ Grassmann variables $\bar\psi_a^i(n)$ if $Q > 0$,
  and the same number of Grassmann variables $\psi_a^i(n)$ if $Q < 0$.
\item For the coefficients $\chi^{Q}\nmmuh$ the situation is the same,
  except that $\psi(n)$ and $\bar\psi(n)$ switch roles.
\item Each term of the expansion of $\e^{\psib Z\psi + \psib Z^\dagger
    \psi + 2am\bar\psi\psi}$ involves as many $\psib(n)$ as $\psi(n)$.
\end{itemize}
The Grassmann integral $\int d\psi(n)d\psib(n)$ extracts the
coefficient of the top--monomial,
\begin{displaymath}
  \int d\psib(n)d\psi(n) \, \prod_{i = 1}^{N_c} \prod_{a = 1}^{N_f}
  \psi^i_a(n) \psib^i_a(n) = 1 \;,
\end{displaymath}
setting all others to zero.  This monomial contains as many $\psib(n)$
as $\psi(n)$.  Hence, in view of the counting above, the contribution
from a configuration $\{Q\npmuh\}$ vanishes unless the following
condition is met:
\begin{equation}\label{dj=0}
  \sum_{\mu = 0}^d Q\npmuh = \sum_{\mu = 0}^d Q\nmmuh \;.
\end{equation}
The physical meaning of this equation is conservation of the baryon
current: the (algebraic) number of baryons ``arriving'' at the site
$n$ (from the links $n - \hmu/2$) must equal the number of baryons
``leaving'' the site (via the links $n + \hmu/2$).

The general structure of the partition function (\ref{contributions})
corresponds to the hadronic correlation function written in terms of
colorless $N_c$--quark currents \cite{Ioffe,Bochum}.

\subsection{Integration over the fermions}

Based on the general considerations above, we perform the integration
over the fermions sector by sector, and present below two particular
cases: the ``vacuum'', i.e.~the sector where the baryonic flux $Q
\npmuh$ vanishes on every link $n + \hmu/2$, and a toy model of a
static baryon on a mesonic background (with $Q\npmuh = 1$ along a time
axis).

In each case, integration over the Grassmann variables yields a purely
bosonic effective action, which depends on the configuration of the
fields $Z$ and $Z^\dagger$.  After computing these effective actions,
we will look for their saddle--point configurations to estimate the
respective partition functions.

Before computing the effective actions in particular cases, we
emphasize the consequences of chiral symmetry, which emerges in the
limit of zero quark mass.  The hypercubic lattice is bipartite, so it
can be split into two sublattices according to the parity of $|n|
\defi \sum\limits_\mu n_\mu$.  Given this splitting, the effective
actions $S(Z, Z^\dagger)$ of all sectors are invariant under the
following global transformation:
\begin{alignat}{2}\label{chiralaction}
  |n|\ \text{even}~: &\quad Z\npmuh \mapsto U_1 Z\npmuh U_2 \;, \quad
  & Z^\dagger\npmuh &\mapsto U_2^\dagger Z^\dagger\npmuh U_1^\dagger
  \;, \\ |n|\ \text{odd}~: &\quad Z\npmuh \mapsto U_2^\dagger Z\npmuh
  U_1^\dagger \;, \quad & Z^\dagger\npmuh &\mapsto U_1 Z^\dagger\npmuh
  U_2 \;, \nonumber
\end{alignat}
for any pair $(U_1,U_2) \in \Gr[N_f]{U} \times \Gr[N_f]{U}$.
Therefore, in each sector, the saddle--point configurations in the
chiral limit form a continous set (namely an ``orbit'') generated by
acting with the chiral symmetry group $\Gr[N_f]{U} \times
\Gr[N_f]{U}$.  As soon as the quark masses are turned on, this
degeneracy disappears, and the saddle points become isolated.
Equations (\ref{chiralaction}) show that the fields $Z,\ 
Z^\dagger\npmuh$ transform differently according to the parity of
$|n|$.  To stress this difference, we give different names to the
fields on different sublattices: the fields living on the ``even''
lattice links will be called $V,\ V^\dagger\npmuh$, while the fields
on the ``odd'' links will be denoted by $W,\ W^\dagger\npmuh$.

\subsubsection{Vacuum action}

For the vacuum sector we have zero baryonic flux $(Q = 0)$ everywhere
on the lattice; the integral over the fermions, being Gaussian, is
then easily done and yields
\begin{eqnarray}
  \mathcal{Z}_{\text{vacuum}} &=& \int d\bar\psi(n) d\psi(n) \,
  \mathcal{Z}_{\text{vacuum}}(\psi, \bar{\psi}) \nonumber \\ &=&
  \int\{\prod_{n,\mu} \alpha_0 \ d\mu(Z, Z^\dagger\npmuh)\} \, \exp(-
  N_c S_{\text{vacuum}}[Z]) \;, \label{vacpartn}
\end{eqnarray}
where the result of the integration has been sent back to the
exponent.  The integration measure in curly brackets will be denoted
by $\mathcal{D}(Z,Z^\dagger)$ in the following.  The factor $N_c$ in
the exponent comes from the color content of the fermions: since the
action $S_Z(\psib,\psi)$ does not couple fermions with different
colors, the Grassmann integral is a product of $N_c$ identical
integrals.  The effective action is
\begin{equation}\label{action-vac}
  S_{\text{vacuum}} = - \sum_n \Tr \ln M(n) + \sum_n \sum_{\mu = 0}^d
  \Tr \ln N\npmuh \;,
\end{equation}
where
\begin{eqnarray}
  M(n) &\defi& 2am\;+\;\sum_{\mu=0}^d \left( Z\npmuh + Z^\dagger\nmmuh
    \right) \;, \\ N\npmuh &\defi& 1 + Z\npmuh Z^\dagger\npmuh \;.
\end{eqnarray}

\subsubsection{Static baryon action}\label{sec:statbar}

By the static baryon we mean the following distribution of baryonic
fluxes over the lattice: $Q \npmuh = 1$ along the links of the ``world
line'' (or ``string'') $n = (t,0,\ldots,0) \in {\mathbb Z}^{1+d}$,
$\hmu = \hat 0$, with $t = 0, \ldots, T-1$; on all other links $Q =
0$.  This distribution satisfies the current conservation law
(\ref{dj=0}) at all sites but the ends $t = 0$ and $t = T$ of the
world line.  There it does, too, if we impose periodic (or
antiperiodic) boundary conditions on the Grassmann fields: for a
lattice of size $T$ in the time direction, we set $\psi( T+1,\vec r) =
\psi(1,\vec r)$, $\bar\psi(T+1,\vec r) = \bar\psi(1, \vec r)$.  We
again write the partition function in the form (\ref{vacpartn}),
\begin{equation}\label{barpartfctn}
  \mathcal{Z}_{\text{baryon}} = \int \mathcal{D} \bar\psi \mathcal{D}
  \psi \, \mathcal{Z}_{\text{baryon}}(\psi, \bar{\psi}) = \int
  \mathcal{D}( Z, Z^\dagger) \, \exp(- N_c S_{\text{baryon}}[Z]) \;.
\end{equation}
The effective action $S_{\text{baryon}}$ contains the ``sea'' term
$S_{\text{vacuum}}[Z]$, plus an extra part coming from the factors
$\chi_1$ along the world line of the baryon.  These factors depend on
the values of the $Z$ field along this line and on the adjacent links,
through the following matrix:
\begin{equation}\label{propagatorworldline}
  \mathsf{G} \defi N({\Ss \frac{1}{2} \hat 0}) M({\Ss 1 \hat 0})^{-1}
  N({\Ss \frac{3}{2} \hat 0}) \cdots N({\Ss (T - \frac{3}{2}) \hat 0})
  M({\Ss(T-1)\hat 0})^{-1} N({\Ss (T - \frac{1}{2}) \hat 0}) M({\Ss T
    \hat 0})^{-1} \;.
\end{equation}
(We use the abbreviation $t\hat 0 \equiv 0 + t\hat 0$ to denote the
sites or links on the world line of the baryon.)  This product of
matrices runs over all sites $n$ on the baryon world line (it is
expressed as a ``quark propagator'' along that line).  In Appendix
\ref{staticbaryon} we show that the effective action takes the form
\begin{equation}
  \label{staticbaryonaction}
  \e^{-N_c\, S_{\text{baryon}}[Z]} = \frac{1}{N_c!} \Big\{ \frac{
    \alpha_0} {\alpha_1} \binom{N_c+N_f-1}{N_c} \Big\}^{-T} \sum_{\hat
    \sigma\in\hat{\mathfrak S}_{N_c}} \mathcal{N}(\hat\sigma)
  \prod_{l=1}^{N_c} \left( \Tr \mathsf{G}^l \right)^{c_l(\hat\sigma)}
  \e^{-N_c\, S_{\text{vacuum}}[Z]} \;.
\end{equation}
In the non--vacuum factor of (\ref{staticbaryonaction}), $\hat\sigma$
runs over all conjugacy classes of the group $\mathfrak{S}_{N_c}$ of
permutations of the set $\{1,...,N_c\}$.  Every representative of the
class $\hat\sigma$ can be decomposed as a product of cycles of various
lengths $l$, such that $c_l(\hat\sigma)$ cycles of length $l$ occur;
thus, each class $\hat\sigma$ is uniquely specified by the sequence
$\{c_l\}$, or equivalently by a Young diagram.  The weight factor
$\mathcal{N}(\hat\sigma)$ is simply the cardinality of the class
$\hat\sigma$, and is given by
\begin{equation}\label{normalizationconstants}
  \mathcal{N}(\hat\sigma) = \frac{N_c!} {\prod\limits_{l=1}^{N_c}
    l^{c_l(\hat\sigma)}c_l(\hat\sigma)! } \;.
\end{equation} 
For the lowest numbers of colors the explicit expressions are
\begin{alignat}{2}
  N_c&=1: \qquad &S_{\text{baryon}} &= S_{\text{vacuum}} - \ln \Tr
  \mathsf{G} + {\rm const} \;, \nonumber \\ \label{baryonN=2} N_c &=2:
  &S_{\text{baryon}} &= S_{\text{vacuum}} - \frac{1}{2} \ln \left((\Tr
    \mathsf{G})^2 + \Tr \mathsf{G}^2 \right) + {\rm const} \;, \\ N_c&=3:
  &S_{\text{baryon}} &= S_{\text{vacuum}} - \frac{1}{3} \ln \left((\Tr
    \mathsf{G})^3 + 3 \Tr \mathsf{G}^2 \Tr \mathsf{G} + 2 \Tr
    \mathsf{G}^3 \right) + {\rm const} \;. \nonumber
\end{alignat}
The constants, which are not given above, make contributions to the
baryon mass, so they need to be taken into account in the final
answer.

In the following section, we look for the saddle--point configurations
of the effective actions $S_{\text{vacuum}}$ and $S_{\text{baryon}}$.

\subsection{Saddle--point equations}\label{saddlepoint}

In the two sectors that we are interested in -- the vacuum and the
static baryon -- we wish to compute, or at least estimate, the
partition functions (\ref{vacpartn}) resp.~(\ref{barpartfctn}).  Since
we are unable to provide an exact answer, we will treat both integrals
in a saddle--point approximation, valid in the limit of a large number
of colors $N_c$.  For both the vacuum and the static baryon, we will
restrict ourselves to a purely classical approximation, which is to
say we will identify the saddle points, evaluate the action functional
on them, and approximate the partition function as $\mathcal{Z} \sim
\e^{-S(Z_{\rm s.p.})}$.  Thus we neglect all loop corrections, which
are of higher order in $1/N_c$.

In the vacuum sector, where $N_c$ appears explicitly as a factor of
$S_{\text{vacuum}}[Z]$, the saddle--point approximation is fully
justified in the large--$N_c$ limit.  The situation is less
transparent in the static--baryon sector (\ref{staticbaryonaction}).
However, for the ansatz made below, the matrix $\mG$ is proportional
to unity: $\mG = \mathsf{g}\mathbb{I}_{N_f}$.  (Note that $\mG$
transforms under the chiral transformation (\ref{chiralaction}) as
$\mG \mapsto U \mG U^{-1}$, so the multiples of unity are fixed points
of this group action.)  If one decides to consider only those
configurations of $Z$ and $Z^\dagger$ for which $\mG$ is scalar, the
static--baryon action (\ref{staticbaryonaction}) simplifies to
\begin{displaymath}
  S_{\rm baryon}[Z] = S_{\rm vacuum}[Z] - \log\mathsf{g} + {\rm
    const} \;,
\end{displaymath}
so the saddle--point expansion is rigorously justified (for large
$N_c$) if the integral is restricted to these configurations.  We will
use it to approximate the full integral.

\section{Vacuum saddle point configurations}\label{vac-config}

The saddle--point analysis for the action functional $S_{\rm vacuum}
[Z]$ has already been carried out in \cite{as,BS,Nagao}, so we are
going to be brief here.  In varying the action (\ref{action-vac}), the
complex matrices $Z$ and $Z^\dagger$ are to be considered as
independent, which leads to two sets of equations.  Variations of
$Z\npmuh$ affect only the blocks $M(n)$ and $N\npmuh$, with the linear
response being
\begin{displaymath}
  \delta S_{\text{vacuum}} = \Tr \left( - M(n)^{-1} + Z^\dagger\npmuh
  N\npmuh^{-1} \right) \delta Z\npmuh \;.
\end{displaymath}
The resulting saddle--point equation reads $M(n)^{-1} = Z^\dagger
\npmuh N\npmuh^{-1}$ or, by taking the inverse on both sides,
\begin{equation}\label{vac-s.p.z}
  M(n) = Z\npmuh + Z^\dagger\npmuh^{-1} \;.
\end{equation}
Similarly, the variation $\delta Z^\dagger\nmnuh$ influences $M(n)$
and $N\nmnuh$, and yields the saddle--point equation $M(n)^{-1} =
N\nmnuh^{-1} Z\nmnuh$, which is equivalent to
\begin{equation}\label{vac-s.p.zbar}
  M(n) = Z^\dagger\nmnuh + Z\nmnuh^{-1} \;.
\end{equation}
As an immediate corollary, we have
\begin{equation}\label{corollary}
  Z\npmuh + Z^\dagger\npmuh^{-1} = Z^\dagger\nmnuh + Z\nmnuh^{-1}
\end{equation}
at every site $n$ and for any pair $\mu, \nu$.

\subsection{Homogeneous vacuum}\label{homogvac}

The simplest possibility for the field $Z,\ Z^\dagger$ is the scalar
ansatz $Z = Z^\dagger =z\mathbb{I}$, with $z$ a spacetime--independent
real number.  The vacuum saddle--point equations (\ref{vac-s.p.z}) and
(\ref{vac-s.p.zbar}) are solved by this ansatz if we put
\begin{equation}\label{z+-}
  z = z_{\pm} = \pm \frac{1}{\sqrt{2d+1}}\sqrt{1+\frac{(am)^2} {2d+1}} -
  \frac{am}{2d+1} \;.
\end{equation}
If $m > 0$, the action $S_{\rm vacuum}$ takes different values on
these solutions.  Expanding it in powers of $(am)$, we get
\begin{equation}\label{valueS}
  S_{\rm vacuum}[z_{\pm}] = L^{d+1} \, N_f \left[ \ln\left( 
      \frac{(2d+2)^d}{(2d+1)^{d+1/2}} \right)\ \mp \ 
    \frac{\sqrt{2d+1}}{d+1}am \right] \;,
\end{equation}
which shows that for a positive quark mass, the configuration $z
\equiv z_+$ minimizes the action.

In the chiral limit ($m = 0$), a continuous set of solutions is
obtained by applying the transformations \eqref{chiralaction} to the
homogeneous configuration $Z = Z^\dagger = z_{\rm vac}\mathbb{I}$ for
\begin{displaymath}
  z_{\rm vac} = (2d+1)^{-1/2} \;.
\end{displaymath}
This vacuum configuration is invariant under the transformations of
the diagonal subgroup $U_1 = U_2\in \Gr[N_f]{U}$ of the chiral
symmetry group, but it maps to a new ``vacuum'' by taking $U_1 = U \in
\Gr[N_f]{U}$, $U_2 = \mathbb{I}_{N_f}$.  By the Goldstone mechanism,
the breaking of the continuous $\Gr[N_f]{U}$ symmetry leads to the
existence of massless modes, namely the mesons, an effective
Lagrangian for which was obtained by expanding $S_{\text{vacuum}}$
near this vacuum in \cite{as,BS}.

If $\pm \mathbb{I} \not= U \in \Gr[N_f]{U}$, the vacua obtained by
translating the homogeneous one by $U$ are staggered, in the sense
that the value of $Z$ depends on the parity of its position.  However,
on adopting the notations $V\npmuh$, $W\nppmuh$ for fields on the even
and odd sublattices, the staggered vacua become homogeneous for each
sublattice:
\begin{equation}\label{homog-vacuum}
  V\npmuh = (2d+1)^{-1/2} U \;, \qquad W\nppmuh = (2d+1)^{-1/2}
  U^\dagger \;.
\end{equation}

In \cite{Nagao}, it was proved that, modulo the $\Gr[N_f]{U}$
degeneracy, the configuration $Z\equiv z_{\rm vac}$ is the
\emph{unique} solution of the vacuum saddle--point equations in the
chiral limit (except in dimension $d = 0$, where the symmetry of the
action is larger).  The proof proceeds by a local argument, showing
that for each site $n$ the $2(d+1)$ saddle--point equations involving
$M(n)$ imply the equality of the matrices $Z\npmuh$, $Z^\dagger\nmmuh$
for all $\mu = 0,\ldots,d$; iteration of this result then trivially
leads to the set of staggered configurations (\ref{homog-vacuum}).
The proof strongly relies on $Z^\dagger$ being the Hermitian conjugate
of $Z$, a constraint which is not mandatory.  By relaxing it, we are
now going to find a plethora of additional solutions of the vacuum
saddle--point equations.

\subsection{Nonhomogeneous vacuum configurations}

By {\it local} considerations, as stated above, the only solutions of
the vacuum saddle--point equations (and the Hermiticity constraint
relating $Z$ to $Z^\dagger$) in the chiral case are homogeneous in
both sublattice fields $V$ and $W$.  However, on a finite lattice, say
with the topology of a $(d+1)$--dimensional torus $L^d \times T$,
there is also a {\it global} aspect to consider: one has to make a
choice of boundary conditions for the various fields.  The simplest
choice are periodic boundary conditions in all directions, but one can
also impose $\theta$--twisted boundary conditions, say along the first
spatial direction $\hat 1$:
\begin{equation}\label{twist}
  V({\scriptstyle n+\frac{\hat\mu}{2}}+L\hat 1) = \e^{\mi\theta}
  V\npmuh \;, \qquad W({\scriptstyle n' + \frac{\hat\mu}{2}} + L\hat
  1) = \e^{-\mi\theta}W\nppmuh \;,
\end{equation}
for all $n$ and $\hat\mu$.  An opposite twist for the fields $V$, $W$
is natural in view of their opposite behavior under the chiral
transformations \eqref{chiralaction}.

Now, accepting these twisted boundary conditions, let us investigate
which configuration will minimize the action \eqref{action-vac}.  A
homogeneous configuration suffers from a ``phase jump'' along a
$d$--dimensional boundary, which is energetically very costly.  A more
reasonable ansatz for a minimum of the action is the following: the
fields $V$, $W$ smoothly rotate their phase, starting from $V,\ 
W\approx z_{\rm vac}$ for $n_1=1$, to $V =\e^{\mi\theta}z_{\rm vac}$,
$W = \e^{- \mi \theta}z_{\rm vac}$ at $n_1 = L$, with a linear phase
evolution in between.  In this way, everywhere in spacetime the
configuration \emph{locally} looks like one of the degenerate
homogeneous vacua.

\subsubsection{Contour deformation}\label{contour}

The above ansatz for $V, W$ is only qualitative.  In order to actually
obtain field configurations that satisfy \emph{both} the twisted
boundary conditions \eqref{twist} \emph{and} the saddle--point
equations (\ref{vac-s.p.z}, \ref{vac-s.p.zbar}), we need to relax the
Hermiticity relation between the fields $Z$ and $Z^\dagger$.

By its construction via the color--flavor transformation, the
integrand $\e^{-S_{\rm vacuum}}$ is to be viewed primarily as a
function of the \emph{real} variables $\{(Z_{ab} + Z^\dagger_{ba})
\npmuh,\ \mi(Z_{ab} - Z^\dagger_{ba}) \npmuh\}$, the total number of
which is $D = 2N_f^2\, (d+1)\, (TL^d)$.  From \cite{Nagao}, this
function for $\theta \notin 2\pi\mathbb{Z}$ has no saddle points on
$\mathbb{R}^D$, but it can be analytically continued into
$\mathbb{C}^D$, where \emph{complex saddle points} may exist.  On such
a saddle point there must exist at least one link $\npmuh$ where the
matrix $Z^\dagger$ differs from the Hermitian conjugate of $Z$.

If a complex saddle point is not ``too far'' from the original contour
of integration, it contributes to the vacuum--sector partition
function, upon deforming the contour of integration so as to reach
that point.

\subsubsection{Vacuum saddle point equations for twisted fields}
\label{vacuum1+1} 

We will demonstrate below the existence of complex saddle points for
the vacuum sector with any $\theta$--twist.  To make things simpler,
we restrict ourselves to a 2--dimensional spacetime, with a twist in
the spatial boundary conditions (we call the time index $t$, the
spatial index $x$).  The above qualitative ansatz for the fields $V$,
$W$ suggests the following symmetries:
\begin{itemize}
\item All fields are scalar, i.e.~at each point $Z$ and $Z^\dagger$
  are multiples of the identity matrix.
\item The fields $V$, $W$ are time--independent.  For each position
  $x$, there are 4 field variables associated with the time--like link
  which we denote by $v_0(x),\ w_0(x),\ v_0^*(x),\ w_0^*(x)$, and 4
  field variables associated with the space direction, which we denote
  by $v_1(x+1/2),\ w_1(x+1/2),\ v_1^*(x + 1/2),\ w_1^*(x+1/2)$.
\end{itemize}
Thus, at each position $x = 0, \ldots, L-1$ we have 8 independent
complex variables.  In the chiral limit $(m = 0)$, the saddle--point
equations \eqref{vac-s.p.z} pertaining to $M(x,t)$ on an even site
$(x,t)$ read
\begin{equation}\label{scalar-vac}
\begin{split}
  v_0(x) + w_0^*(x) + v_1(x+1/2) + w_1^*(x-1/2) &= v_0(x)+1/v_0^*(x)\\ 
  &= w_0^*(x)+1/w_0(x)\\ &= v_1(x+1/2)+1/v_1^*(x+1/2)\\ &=
  w_1^*(x-1/2)+1/w_1(x-1/2) \;.
\end{split}
\end{equation}
The equations \eqref{vac-s.p.zbar} pertaining to $M(x,t+1)$ are
obtained by interchanging $v\leftrightarrow w$, $v^*\leftrightarrow
w^*$.  For a finite quark mass, $2am$ is to be added to the left--hand
side.
\begin{itemize}
\item The two first equations, together with their $v\leftrightarrow
  w$ exchange analogs, allow us one more simplification.  Indeed, they
  imply the identities $v_0(x) = w_0^*(x)$, $v_0^*(x) = w_0(x)$ (the
  alternative possibility, $v_0(x) = 1/w_0(x)$ and $v_0^*(x) = 1 /
  w_0^*(x)$, is incompatible with other relations that need to be
  satisfied).  So there remain only 6 complex variables for each $x$.
\end{itemize}
In the next section, we will provide approximate solutions for
Eq.~\eqref{scalar-vac} together with their $v\leftrightarrow w$
partners and assuming the above symmetries.

\subsubsection{Linearized problem}\label{linear}

To solve (at least approximately) the above equations, we will use the
fact that we expect the fields to be locally close to one of the
configurations \eqref{homog-vacuum}.  We can then expand the
saddle--point equations to first order in the perturbations from that
configuration, and solve the linear problem.  We start by expanding
the fields around the real positive vacuum $V,\ W = \mathbb{I} /
\sqrt{3}$:
\begin{align}
  v_0(x) &= \isq3(1+\delta v_0(x)) \;, &v_0^*(x) &= \isq3 (1+\delta
  v_0^*(x)) \;, \nonumber \\ v_1(x+1/2) &= \isq3 (1+\delta v_1(x+1/2))
  \;, &v_1^*(x+1/2) &= \isq3(1+\delta v_1^*(x+1/2)) \;, \nonumber \\ 
  w_1(x+1/2) &= \isq3(1+\delta w_1(x+1/2)) \;, &w^*_1 (x+1/2) &=
  \isq3(1+\delta w^*_1(x+1/2)) \;. \nonumber
\end{align}
After inserting these expressions into \eqref{scalar-vac} and
expanding to linear order, we obtain a ``transfer matrix
representation'' of these equations, i.e.~a linear equation relating
the vector of deviations of the spatial components of the fields
$\{\delta v_1,\delta v_1^*,\delta w_1,\delta w_1^*\}$ at position
$x+1/2$, to the same vector at position $x-1/2$.  The structure of the
$4\times 4$ transfer matrix allows to decompose it into two $2\times
2$ matrices, upon considering at each point the vectors
\begin{equation}\label{drdi}
  \dr = \binom{\delta v_1+\delta w_1}{\delta v_1^*+\delta w_1^*} \;,
  \qquad \di = \binom{\delta v_1-\delta w_1}{\delta v_1^*-\delta
    w_1^*} \;.
\end{equation}
In terms of these two vectors, the linearized equations read
\begin{align}\label{linear-vacuum}
  \dr (x+1/2)& =\begin{pmatrix}-6&1\\-1&0\end{pmatrix}\dr(x-1/2) \defi
  \mathsf{T}_r\ \dr(x-1/2) \;, \\ \di (x+1/2) &= \begin{pmatrix} 3/2
    &-1/2 \\ 1/2 &1/2 \end{pmatrix} \di(x-1/2) \defi \mathsf{T}_i\ 
  \di(x-1/2) \;.
\end{align}
Similarly, the deviations of the temporal components $\delta v_0(x)$,
$\delta v_0^*(x)$ are determined by the variations at $x-1/2$:
\begin{equation}\label{linearv_0}
  \binom{\delta v_0(x)}{\delta v_0^*(x)} =
  \begin{pmatrix}3/4&-1/4\\3/4&-1/4\end{pmatrix}\dr(x-1/2)
  + \begin{pmatrix}3/8&-1/8\\-3/8&1/8\end{pmatrix}\di(x-1/2) \;.
\end{equation}
Thus, the transfer matrix allows to express the linear variation of
all fields by the deviations at position $1/2$.

To make the $x$--dependence more explicit, we seek to diagonalize the
transfer matrices $\mathsf{T}_r$ and $\mathsf{T}_i$.  The first
transfer matrix $\mathsf{T}_r$ has the eigenvalues
\begin{equation}\label{eigenbasis-real}
  -\e^{\pm\lambda}\defi -(3\pm 2\sqrt{2}),\quad\mbox{associated to the
    vectors}\ \mathsf{R}_\pm\defi\binom{1}{\e^{\mp\lambda}}.
\end{equation}
The second transfer matrix $\mathsf{T}_i$ cannot be diagonalized but
only put in Jordan normal form.  Indeed, it acts on the vectors
\begin{equation}
  \mathsf{I}_+\defi\binom{1}{1},\ \ \ \mathsf{I}_-\defi\binom{1}{-1}
\end{equation}
as
\begin{displaymath}
  \mathsf{T}_i\ \mathsf{I}_+ = \mathsf{I}_+\,+\,\mathsf{I}_- \;,
  \mathsf{T}_i\ \mathsf{I}_- = \mathsf{I}_- \;.
\end{displaymath}
Therefore, an initial deviation
\begin{displaymath}
  \dr (1/2) = c_{+r}\ \mathsf{R}_+ +c_{-r}\ \mathsf{R}_- \;, \qquad
  \di(1/2)= c_{+i}\ \mathsf{I}_+ +c_{-i}\ \mathsf{I}_-
\end{displaymath}
propagates through the transfer matrix as follows:
\begin{equation}\label{result-linear-vac}
\begin{split}
  \dr (x+1/2)&= c_{+r}(-\e^\lambda)^x\ \mathsf{R}_+ + c_{-r}
  (-\e^{-\lambda})^x\ \mathsf{R}_- \;,\\ \di (x+1/2) &= c_{+i}\ 
  \mathsf{I}_+ + (c_{-i}+x\; c_{+i})\ \mathsf{I}_- \;.
\end{split}
\end{equation}
This linear evolution is only valid as long as the deviations from
$1/\sqrt{3}$ are small compared to unity.  This cannot be the case
uniformly for our twisted ansatz, where the fields near $x = L$ take
values close to $\e^{\pm\mi\theta}/\sqrt{3}$.  Still, the fact that
$\di(x+1/2)$ depends \emph{linearly} on the position is encouraging:
this is exactly the behavior we expect for the \emph{phases} of the
fields in the ansatz.

The linearization of the saddle--point equations can actually be
performed near \emph{any} of the degenerate family of vacua
(\ref{homog-vacuum}).  Linearizing the equations in the vicinity of a
vacuum $v = \e^{\pm\mi\varphi}/\sq3$, we obtain for the deviations the
same transfer matrix as before.  We can therefore construct local
solutions near various $\varphi$--vacua using
\eqref{result-linear-vac}, and glue them together to obtain a global,
``rotating'' solution.  An equivalent procedure is to exponentiate the
deviations,
\begin{equation}\label{scalar-Ansatz-nlin}
  v_0(x) = \isq3 \exp\{\delta v_0(x)\} \;,
\end{equation}
etc., and extend the equations \eqref{result-linear-vac} for $\di(x +
1/2)$ to a larger domain of validity.  This we do as follows.

First of all, the $\dr$--part of the deviations grows exponentially,
and is staggered with respect to $x$.  Our ansatz excludes both
features, so we simply set $c_{+r} = c_{-r} = 0$.  Next, we note that
the $\di$--deviations depend on two coefficients, $c_{+i}$ and
$c_{-i}$.  According to \eqref{scalar-Ansatz-nlin}, their real parts
describe the moduli of the fields, and the imaginary parts the phases.
In our ansatz, we expect the moduli of the fields to be constant and
close to $1/\sq3$ (a linear growth would be incompatible with the
boundary conditions).  Therefore, we set $\re c_{+i} = 0$.

The other coefficient $c_{-i}$ causes a global shift of the fields,
which can be interpreted as a ``generalized chiral rotation'': the
generalization consists in taking in Eqs.~\eqref{chiralaction} for
$U_1$ and $U_2$ any \emph{invertible complex matrix}, and replacing
$U_1^\dagger$ and $U_2^\dagger$ by $U_1^{-1}$ and $U_2^{-1}$.  One
easily checks that the effective action is invariant under this
$\Gr[N_f]{GL}\times \Gr[N_f]{GL}$ extension of the chiral symmetry
group.  The complex extension appears since we have relaxed the
Hermiticity condition.  The parameter $c_{-i}$ is then seen to
parametrize the $\mathbb{C}^{\times}$--manifold of scalar complex
homogeneous vacua.

The remaining coefficient, which we abbreviate to $\alpha \defi \im
c_{+i}$, is responsible for a nonhomogeneous solution.  As we
explained above, deviations $\di(x)\sim \mi\alpha x$ of order ${\cal
  O}(1)$ can be rescaled to deviations of order ${\cal O}(\alpha)$ by
changing the reference vacuum $\e^{\pm\mi\varphi}/\sq3$.  As a result,
the fields built from (\ref{scalar-Ansatz-nlin},
\ref{result-linear-vac}) with $c_{\pm r} = 0$, $\re c_{+i} = 0$
satisfy the saddle--point equations up to order ${\cal O}(\alpha^2)$
uniformly in $x$.

One can add terms of higher order in $\alpha$ to the exponent, so as
to kill the higher--order terms in the expansion of
\eqref{scalar-vac}.  By iterating the procedure, one obtains for the
fields a series expansion in powers of $\alpha$, such that the
saddle--point equations are satisfied to any order.  We conjecture
that this expansion can be (re)summed, at least in a certain domain in
$\alpha$, thereby yielding an exact solution of \eqref{scalar-vac}.
The solution up to order $\alpha^2$ is
\begin{equation}\label{result-nlin-vac}
\begin{split}
  v_0(x)&=\frac{\e^{\mi\alpha x}}{\sq3}\ \e^{c_{-i}-\mi\alpha/2
    +5\alpha^2/8} \;, \\ v_1(x+1/2)&=\frac{\e^{\mi\alpha x}}{\sq3}\ 
  \e^{c_{-i}+\mi\alpha -\alpha^2/8} \;, \\ v_1^*(x+1/2) &=
  \frac{\e^{-\mi\alpha x}}{\sq3}\ \e^{-c_{-i}+\mi\alpha -\alpha^2/8}.
\end{split}
\end{equation}
The expressions for the fields $w$, $w^*$ are obtained by replacing
$\alpha\to -\alpha$, $c_{-i}\to -c_{-i}$.

The coefficient $\alpha$ parametrizes the slope of the phase with
respect to $x$, and it must be tuned according to the boundary
conditions:
\begin{equation}\label{alpha}
  \alpha L = \theta + 2\pi Q_w \;,
\end{equation}
where $Q_w$ is some integer.  For a finite twist $\theta$, $\alpha$ can
be chosen small only in the large--volume limit $L\gg 1$, in which
case $\alpha$ can take several values labeled by the integers $Q_w\ll
L$.

\subsubsection{Topologically nontrivial configurations}

We now return to the original problem with periodic boundary
conditions ($\theta = 0$).  We have shown that there exist nontrivial
solutions, for which the fields are position--dependent, with their
phases rotating $Q_w$ times when the position $x$ goes from $0$ to
$L$.  The integer $Q_w$ can be called the \emph{winding number} of the
configuration.  We can associate a winding number to a (discrete)
configuration because the phases of the fields $v,\ w$ are varying
\emph{smoothly} with position.  More generally, when the lattice has
the topology of a $(1+d)$--dimensional torus, one can associate to any
smooth scalar configuration a set of winding numbers $\{Q_{w, \hat\mu
  }\}$, each number specifying the number of times $\arg(v)$ rotates
between the positions $\npnuh$ and $({\scriptstyle n + L \hat\mu +
  \frac{\hat\nu} {2}})$.

Using Newton's algorithm, we have searched for numerical solutions of
the vacuum saddle--point equations \eqref{scalar-vac}, starting from
trial configurations with $Q_w = 1$, $Q_w = 2$.  We plot some results
in Fig.~\ref{vac120-Q1m000}.  These plots are very well described by
our approximation \eqref{result-nlin-vac}, including the ${\cal O}
(\alpha^2)$ corrections.

\subsubsection{Nontrivial vacua for finite quark mass}
\label{linear-m}

So far we have constructed nontrivial vacua only in the chiral limit,
where a continuum of homogeneous vacuum configurations exists.  What
happens to these nontrivial vacua when the chiral symmetry is broken
explicitly by switching on the quark mass?

Recall that for $m \not= 0$ there remain only two homogeneous scalar
vacua \eqref{z+-} out of the former ${\rm U}(1)$ continuum, with one
of them ($z_{\rm vac} = z_+$) being an absolute minimum of the action.
One can again study the linearized saddle--point equations near this
solution.  Unlike before, the results will apply only locally, since
we cannot use the trick of rescaling the homogeneous reference vacuum
any more.

We have performed numerical searches for $(1+1)$--dimensional
topologically nontrivial vacua, assuming the same symmetries as before
(we took scalar, time--independent, smooth\-ly varying fields).  The
saddle--point equations are modified by the addition of the quark mass
term $2am$ to the left--hand side of Eqs.~\eqref{scalar-vac}.  The
outcome of these calculations (see below) can be understood in large
part by analytical reasoning, as follows.

To linearize the saddle--point equations around the point $z_+$, we
set
\begin{equation}\label{m-nlin}
  v_0(x) = z_+ \e^{\delta v_0(x)} \;, \quad {\rm etc.}
\end{equation}
As in the chiral case, the $4\times 4$ transfer matrix splits into two
$2\times 2$ matrices that apply to the vectors $\dr$, $\di$ defined in
Eq.~\eqref{drdi}.  These matrices can be written for an arbitrary
value of $am$, using the exact expression \eqref{z+-} for $z_+(am)$
(we only consider the case $d = 1$).  Using the same notations as
above, they are
\begin{equation}
\begin{split}
  \mathsf{T}_r(am) &= \frac{1}{1-z_+^4}
  \begin{pmatrix}-z_+^{-2}-2-z_+^2&2z_+^2+2z_+^4\\ -2z_+^2-2z_+^4 &
    -z_+^2+2z_+^4+3z_+^6\end{pmatrix} \;, \\ \mathsf{T}_i(am) &=
  \frac{1}{1-z_+^4}\begin{pmatrix}z_+^{-2}-2+z_+^2&2z_+^2-2z_+^4\\ 
    -2z_+^2+2z_+^4 & z_+^2+2z_+^4-3z_+^6\end{pmatrix} \;,
\end{split}
\end{equation}
and the deviations on links pointing in the time direction propagate
as
\begin{equation}
  \binom{\delta v_0(x)}{\delta v_0^*(x)} = \frac{1}{2(1-z_+^2)}
  \begin{pmatrix} 1&-z_+^2 \\ 1&-z_+^2\end{pmatrix} \dr(x-1/2)
  +\frac{1}{2(1+z_+^2)} \begin{pmatrix}1&-z_+^2 \\ -1&z_+^2
  \end{pmatrix}\di(x-1/2) \;.
\end{equation}
One easily checks that both transfer matrices have the property $\det
\mathsf{T}_{r / i}(am) = 1$.  

We expand both matrices and their spectra in powers of $am$, since we
are interested in the case of a small quark mass.  In the massless
limit the matrix $\mathsf{T}_r(0)$ is hyperbolic, with negative real
eigenvalues that are well separated from each other (one expanding,
the other contracting).  Thus, a perturbation of order ${\cal O}(am)$
is still diagonalizable, with eigenvalues and eigenspaces shifted by
that same order ${\cal O}(am)$.  We will keep calling the eigenvalues
$-\e^{\pm\lambda}$, with the expansion
\begin{equation}\label{lambda-m}
  \lambda(am) = \ln(3+2\sqrt{2})+am/\sqrt{6}+{\cal O}((am)^2) \;.
\end{equation} 
On the other hand, $\mathsf{T}_i(0)$ was nondiagonalizable with
eigenvalue $+1$, so a perturbation can change its qualitative
features.  For any positive $am$, $\mathsf{T}_i(am)$ becomes
diagonalizable, with real positive eigenvalues $\e^{\pm\gamma(am)}$
associated to eigenvectors $\binom{1}{e^{\pm\kappa_1(am)}}$.  To
express the deviations of the field $v_0(x)$, we use the coefficients
\begin{displaymath}
  \e^{\kappa_{\pm 0}} \defi \frac{1+z_+^2\e^{\pm\kappa_1}}{1+z_+^2}
  \;.
\end{displaymath}
For small $am$, these data have the following expansions:
\begin{equation}\label{gamma-m}
\begin{split} 
  \gamma(am)&=\frac{2}{3^{1/4}}(am)^{1/2}\ +\ \frac{5}{3^{7/4}}
  (am)^{3/2}\ +\ {\cal O}((am)^{5/2}) \;,\\ \kappa_1(am) &=
  -\frac{4}{3^{1/4}}(am)^{1/2}\ -\frac{4}{3^{7/4}}(am)^{3/2}\ +\ {\cal
    O}((am)^{5/2}) \;,\\ \kappa_{\pm 0}(am) &= \mp\frac{1}{3^{1/4}}
  (am)^{1/2} + \frac{\sqrt{3}}{2}am \pm \frac{1}{2 \times 3^{7/4}}
  (am)^{3/2} + {\cal O}((am)^2) \;.
\end{split}
\end{equation} 
For a finite mass $am$, one expects the linear approximation to be
valid only for small deviations.  However, the error introduced in the
saddle--point equations by the linear approximation is at most of
order ${\cal O}(am|\delta z|)$, so it remains small if the mass is
small.

The numerical solution we obtained for a mass $am=0.01$ and winding
number $Q_w = 1$ (see Figs.~\ref{vac120-Q1m005},
\ref{vac120-Q1m005log}) suggest that the vectors $\dr$ are negligible
in a large domain of $x$ around the point $x=0$ where fields are close
to $z_+$.  This indicates that the coefficients $c_{\pm r}$ vanish,
like in the chiral case.  The fields will therefore depend on two
complex parameters $\eps_{\pm}$:
\begin{equation}\label{result-m-vac}
\begin{split} 
  \delta v_1(x+1/2)&=\eps_{+} \e^{\gamma x}+\eps_{-}\e^{-\gamma x} \;,
  \\ \delta v_1^*(x+1/2) &= -\eps_{+} \e^{\kappa_1+\gamma x}-\eps_{-}
  \e^{-\kappa_1-\gamma x} \;, \\ \delta v_0(x+1)&=\eps_{+}
  \e^{\kappa_{+0}+\gamma x}+\eps_{-} \e^{\kappa_{-0}-\gamma x} \;.
\end{split}
\end{equation} 
As opposed to the chiral case, both coefficients $\eps_{\pm}$ are
complex, so that both the phases and the moduli of the fields vary
with $x$.  This ansatz fits the numerical solution even when the
deviations from $z_+$ become of order ${\cal O}(1)$, which is quite
surprising.  However, it is unable to reproduce the zone where the
fields cross the negative real axis (near $x = L/2$).  For a quark
mass $am = 0.01$ the values of the various exponents are
\begin{equation}\label{datam005}
  \gamma = 0.1527 \;, \qquad \kappa_1 = -0.3045 \;, \quad \kappa_{+0}
  = -0.0673 \;, \qquad \kappa_{-0} = 0.0845 \;.
\end{equation}
These values are used in the fits to the numerical solution shown in
Fig.~\ref{vac120-Q1m005log}.

\section{Static baryon saddle point equations}\label{sec:speqb}

The effective action $S_{\text{baryon}}$ for the static--baryon
sector, Eq.~(\ref{staticbaryonaction}), contains a ``string'' term in
addition to the ``sea'' term $S_{\text{vacuum}}$.  While the sea term
depends on every one of the matrices $Z\npmuh$, the string term
involves only those matrices $Z$ and $Z^\dagger$ that are situated in
the near vicinity of the string.  More precisely, what enters into the
baryon world line propagator, $\mathsf G$, are the matrices $N({\Ss (t
  + \frac{1}{2}) \hat 0})$ and $M({\Ss t\hat 0})$.  Of these, the
former depend only on $Z$ and $Z^\dagger$ along the string, whereas
the latter also involve the matrices $Z({\Ss t\hat 0 + \hmu/2})$ and
$Z^\dagger({\Ss t\hat 0 - \hmu/2})$.  The $Z$ field on the remaining
links (away from the string) appears only in $S_{ \text{vacuum}}$, so
the variation with respect to these matrices yields the same equations
(\ref{vac-s.p.z}) and (\ref{vac-s.p.zbar}) as in the vacuum sector.

For simplicity let us consider the two particular cases $N_c = 2$ and
$N_c = 3$, using the expressions (\ref{baryonN=2}) for the effective
action.  The most general variation yields
\begin{equation}\label{dL}
  \begin{split}
    N_c = 2 &: ~ \delta S_{\text{baryon}} = \delta S_{\text{vacuum}}
    - \frac{\Tr(\mG \delta \mG) + \Tr(\mG) \Tr (\delta \mG)} {\Tr
      (\mG^2) + (\Tr \mG)^2} \;, \\ 
    N_c = 3 &: ~ \delta S_{\text{baryon}} = \delta S_{\text{vacuum}}
    - \frac{(\Tr \mG)^2 \Tr \delta \mG + 2 \Tr \mG \Tr(\mG \delta \mG) +
      \Tr(\mG^2) \Tr (\delta \mG) + 2 \Tr (\mG^2 \delta \mG)} {(\Tr
      \mG)^3 + 3 \Tr \mG\, \Tr (\mG^2) + 2 \Tr \mG^3} . \nonumber
  \end{split}
\end{equation}
We then work out how the various traces of powers of $\mathsf G$
respond to variations of each matrix $Z$ and $Z^\dagger$ entering in
the definition of $\mG$.  For instance, variations of $Z({\Ss t\hat 0
  + \hmu/2})$ with $\mu\neq 0$ affect only the matrix $M({\Ss t\hat
  0})$, whereas varying $Z({\Ss (t+1/2)\hat 0})$ affects both $M({\Ss
  t\hat 0})$ and $N({\Ss (t+1/2)\hat 0})$.  These computations are
simplified by the use of cyclicity properties: given any decomposition
$\mG = \mG_1\mG_2$, we may replace $\mG$ in the static--baryon action
by the matrix $\tilde\mG = \mG_2 \mG_1$, as $\mG$ always appears
under a trace.

We provide detailed calculations for the variation with respect to $\zeta
\equiv Z({\Ss (t + 1/2)\hat 0})$.  The modified factors of $\mG$ in
this case are $M( {\Ss t \hat 0})^{-1}$ and $N({\Ss (t+1/2)\hat 0})$,
and the modified matrix $\mG$ reads
\begin{displaymath}
  \mG + \delta\mG = \cdots M({\Ss t\hat 0})^{-1} \big\{ 1 + \delta
  \zeta \big( - M({\Ss t\hat 0})^{-1} + \zeta^\dagger N ({\Ss (t +
    1/2)\hat 0})^{-1} \big) \big\} N({\Ss (t+1/2)\hat 0}) \cdots \;.
\end{displaymath}
It is now natural to conjugate $\mG + \delta \mG$ into
\begin{displaymath}
  T \left( \mG + \delta\mG \right) T^{-1} = \tilde\mG + \delta\zeta
  \big( -M({\Ss t\hat 0})^{-1} + \zeta^\dagger N({\Ss (t+1/2)\hat
    0})^{-1} \big) \tilde\mG \;,
\end{displaymath}
where $\tilde\mG \defi N({\Ss (t+1/2)\hat 0}) M({\Ss (t+1)\hat
  0})^{-1} \cdots N({\Ss (t-1/2)\hat 0}) M({\Ss t\hat 0})^{-1}$.
The saddle--point equation that follows from varying $Z({\Ss (t +
  1/2)\hat 0})$ then takes the succinct form
\begin{displaymath}
  \delta Z({\Ss (t+1/2)\hat 0}) : \quad 0 = \big( - M({\Ss t\hat
    0})^{-1} + Z^\dagger({\Ss (t+1/2)\hat 0}) N({\Ss (t+1/2)\hat
    0})^{-1} \big) \cdot \big(\mathbb{I}_{N_f} - F_{N_c}(\tilde\mG)
  \big) \;,
\end{displaymath}
with the case--by--case definition of the matrix--valued function
$F_{N_c}(G)$ being
\begin{eqnarray*}
  N_c = 2 &:& \quad F_2(G) = \frac{G^2 + G \Tr G }{\Tr (G^2) + (\Tr
    G)^2} \;, \\ N_c = 3 &:& \quad F_3(G) = \frac{G (\Tr G)^2 + 2 G^2
    \Tr G + G \Tr (G^2) + 2 G^3} {(\Tr G)^3 + 3 \Tr G \, \Tr (G^2) + 2
    \Tr (G^3)} \;.
\end{eqnarray*}
The saddle--point equation obtained by varying $Z^\dagger ({\Ss (t -
  1/2)\hat 0})$ is similar.  It is best expressed in terms of the
matrix $\tilde\mG' = M({\Ss t\hat 0})^{-1} N({\Ss (t+1/2)\hat 0})
\cdots M({\Ss (t-1)\hat 0})^{-1} N({\Ss (t-1/2)\hat 0})$:
\begin{displaymath}\label{varZ_0dagger}
  0 = \big( \mathbb{I}_{ N_f} - F_{N_c}(\tilde\mG')\big) \cdot \big( -
  M({\Ss t\hat 0}) ^{-1} + N({\Ss (t-1/2)\hat 0})^{-1}
  Z({\Ss (t-1/2)\hat 0}) \big) \;,
\end{displaymath}
with the same definitions for $F_{N_c}$ as above.  The term
multiplying the unit matrix $\mathbb{I}_{N_f}$ stems from $\delta
S_{\rm vacuum}$.  Note that this term factors out in both of the above
variations.

To get an idea of the matrix $\mathbb{I}_{N_f}-F_{N_c}(G)$, we compute
it in the vacuum configuration $Z\equiv z_{\rm vac} \mathbb{I}_{N_f}$.
In this case we have $\mG = \tilde\mG = \tilde\mG' \propto \mathbb{I}
_{N_f}$.  We then notice that $F_{N_c}( \alpha \mathbb{I}_{N_f}) =
N_f^{-1} \, \mathbb{I}_{N_f}$ for any number of colors and any $\alpha
\neq 0$.  Therefore, on the vacuum configuration, we get $
\mathbb{I}_{N_f} - F_{N_c}(\mG) = (1 - N_f^{-1}) \mathbb{I}_{N_f}$,
which is invertible as soon as $N_f > 1$.  More generally, this
equation holds as long as $\mG$ is a multiple of the unit matrix,
which is a property of the inhomogeneous scalar ansatz we will make in
the next section.

Clearly, as long as the matrix $\mathbb{I}_{N_f} - F_{N_c}
(\mG)$ remains non--singular, the saddle-point equations due to
varying $Z({\Ss (t+1/2)\hat 0})$ and $Z^\dagger({\Ss (t+1/2)\hat 0})$
are identical to those in the vacuum sector, Eqs.~(\ref{vac-s.p.z})
and (\ref{vac-s.p.zbar}).  As was said earlier, this is also the case
for the equations due to varying all matrices $Z$ and $Z^\dagger$ not
involved in the matrix $\mG$, i.e.~those away from the string.  The
only difference to the vacuum equations comes from the matrices on the
links \emph{adjacent} to the string, namely $Z({\Ss t\hat 0+\hmu/2})$,
$Z^\dagger({\Ss t\hat 0 -\hmu/2})$ for the directions $\mu = 1,
\ldots, d$.  These matrices are contained only in some $M^{-1}$ factor
of $\mG$, and their variations give the following saddle--point
equations:
\begin{alignat}{2}
  \delta Z({\Ss t\hat 0 +\hmu/2}) &: \quad -M({\Ss t\hat 0})^{-1} +
  Z^\dagger({\Ss t\hat 0 +\hmu/2}) N({\Ss t\hat 0 +\hmu/2})^{-1} &=&
  -M({\Ss t\hat 0})^{-1} \, F_{N_c}(\tilde\mG) \;, \label{deltaZhmu}
  \\ \delta Z^\dagger({\Ss t\hat 0 -\hmu/2}) &: \quad - M({\Ss t\hat
    0})^{-1} + N^{-1}({\Ss t\hat 0 -\hmu/2}) \, Z({\Ss t\hat 0
    -\hmu/2}) &=& -F_{N_c}(\tilde\mG') \, M({\Ss t\hat 0})^{-1} \;,
    \label{deltaZ*hmu}
\end{alignat}
where the matrices $\tilde\mG$, $\tilde\mG'$ are the same as before.
These equations represent the only obstruction that prevents the
vacuum configuration $Z = Z^\dagger\equiv z_{\rm vac} \mathbb{I}$ from
being also a saddle point of the static--baryon sector.

\subsection{Configurations for the static baryon in $1+1$ dimensions} 
\label{sec:one+one}

In this section we present approximate solutions of the saddle--point
equations in the static baryon sector for the simplest nontrivial
case, which is the two-dimensional Euclidean square lattice ($d = 1$).
We use the same notations and assume the same symmetries as in Section
\ref{vacuum1+1}, so at each position $x$ there are 8 independent
complex scalar variables.

The baryonic string is placed on the Euclidean time axis at position
$x = 0$.  The equations to solve are the vacuum saddle--point
equations (\ref{scalar-vac}) off the string $(x \not= 0)$, and the
modified equations
\begin{equation}\label{scalar-string}
\begin{split}
  v_0(0) + w_0^*(0) + v_1(1/2) + w_1^*(-1/2)+ 2am &=
  v_0(0)+1/v_0^*(0)\\ &= w_0^*(0)+1/w_0(0)\\ &=
  (1-N_f^{-1})\{v_1(1/2)+1/v_1^*(1/2)\}\\ &=
  (1-N_f^{-1})\{w_1^*(-1/2)+1/w_1(-1/2)\}
\end{split}
\end{equation}
on the string, together with the equations obtained by exchanging $v
\leftrightarrow w$, $v^* \leftrightarrow w^*$.  As in the vacuum
sector, these equations imply the identifications $v_0(x) = w_0^*(x)$
and $w_0(x) = v_0^*(x)$ for all $x$.

\subsubsection{Physical requirements}

Recall from Section \ref{homogvac} that demanding $Z^\dagger$ to be
the Hermitian conjugate of $Z$ (and assuming the vacuum saddle--point
equations) leads to a homogeneous configuration, where the fields are
constant on each sublattice.  Such a homogeneous configuration
\emph{cannot} satisfy the last two of the equations
\eqref{scalar-string}.  To get a solution, we must relax the
Hermiticity condition (cf.~Section \ref{contour}), and consider the
fields $Z$ and $Z^\dagger$ as independent variables.

We want the baryon to be a \emph{localized} object, in the sense that
a baryonic saddle--point configuration should differ from a vacuum
configuration only in some neighborhood of the baryon world line.  The
baryon can then be interpreted as a spatially localized excitation of
this vacuum.  A priori, baryon excitations may exist on top of each of
the vacua described in Section \ref{vac-config}.

In Eqs.~\eqref{scalar-string}, the number of flavors $N_f$ enters just
as a parameter, so one can extend the equations to any real value of
$N_f$.  In the limit $N_f = \infty$, we recover the vacuum
saddle--point equations.  We can therefore obtain a solution of the
baryon saddle--point equations by starting from a given vacuum
configuration (at $N_f = \infty$), and deforming the configuration by
continuous variation of $N_f$ down to its physical value (say $N_f =
2$).  Any baryon configuration obtained in this way carries the same
topological charge as the vacuum it is associated to.

\subsubsection{Topologically trivial sector $Q_w = 0$, chiral limit}
\label{baryon-chiral}

In the sector with zero winding number, we numerically found a unique
solution (see Fig.~\ref{bar120-Q0m000}) asymptotic to the homogeneous
vacuum $z_{\rm vac}$, i.e.~satisfying the asymptotic condition
\begin{displaymath}
  v_\mu,\ v_\mu^*,\ w_\mu,\ w_\mu^*\stackrel{|x|\to \infty}
  {\longrightarrow} z_{\rm vac} = 1/\sqrt{3} \;.
\end{displaymath}
All the fields of this configuration are real and time--independent.
Various components coincide pairwise or in quadruples:
\begin{displaymath}
  v_0 = w_0 = v_0^* = w_0^* \equiv z_0 \;, \quad v_1 = w_1 \equiv z_1
  \;, \quad v_1^* = w_1^* \equiv z_1^* \;.
\end{displaymath}
One easily checks that the saddle--point equations are invariant under
the following transformation:
\begin{equation}\label{x->-x}
  z_1(x+1/2) \leftrightarrow z_1^*(-x-1/2) \;, \qquad z_0(x)
  \leftrightarrow z_0(-x) \;,
\end{equation}
which represents a reflection at the baryon world line.  The solution
found numerically is invariant under this transformation, and we
believe the same to be true for the exact solution (or else we would
get a second solution by reflection).  We can therefore restrict our
study to the domain $0\leq x$.

Our numerics show an exponential convergence of all the fields towards
$z_{\rm vac}$ as we depart from the string (see
Fig.~\ref{bar120-Q0m000}), and the signs of the deviations alternate
with $x$.  This phenomenon can be explained by the linearized
saddle--point equations studied in Section \ref{linear}.  The linear
theory indeed applies if the fields are close to $z_{\rm vac}$, which
is the case for large enough $x$.  Eqs.~(\ref{linearv_0},
\ref{result-linear-vac}), together with the physical condition that
the deviations decay as $x\to\infty$, require $c_{+r} = 0$.  We thus
get the ansatz
\begin{equation}\label{ans-z}
\begin{split}
  z_0(x) &= \isq3(1+c_{-r}\, \sqrt{2} \, \e^\lambda \,
  (-\e^{-\lambda})^x) \;, \\ z_1(x+1/2) &= \isq3(1+c_{-r}\,
  (-\e^{-\lambda})^x) \;, \\ z_1^*(x+1/2) &= \isq3(1+c_{-r}\,
  \e^\lambda\,(-\e^{-\lambda})^x) \;,
\end{split}
\end{equation}
with $\e^\lambda = 3+2\sqrt{2}$ as before.  According to this linear
approximation, the fields oscillate around the asymptotic value
$z_{\rm vac}$, and the amplitude of the oscillations is controlled by
a unique coefficient, which we denote by $C_1 \defi c_{-r}$.

The results \eqref{ans-z} fit the numerical configuration not only far
from the string (where this is expected), but even down to the baryon
string, where the fields deviate significantly from $z_{\rm vac}$.
More precisely, the ansatz fits $z_1$, $z_1^*$ for all $x$, whereas
$z_0$ departs from it only at $x = 0$.  The value $z_0(0) \defi
\frac{1}{\sqrt 3} \left(1 + \tilde C_0\right)$ together with the
parameter $C_1$ can be computed using the (nonlinear) saddle--point
equations on the string (\ref{scalar-string}) and the reflection
symmetry \eqref{x->-x}.  We obtain two equations:
\begin{align*}
  \tilde C_0^2 + 2 C_1 \tilde C_0 + 2 C_1 + 4 \tilde C_0 &= 0 \;,\\ 
  3\e^\lambda C_1^2 + 4\e^\lambda\tilde C_0C_1 + 4\tilde C_0 + C_1
  (3+7\e^\lambda) + 4 &= 0 \;.
\end{align*}
The equations have four pairs of solutions, two real ones and two
complex ones, conjugate to one another.  The physical solution (which
deforms to $C_1 = \tilde C_0 = 0$ as we vary $N_f$ from $2$ to
$\infty$) is $C_1 = -0.0971$ and $\tilde C_0 = 0.0504$, giving
\begin{equation}
  \sqrt{2} \e^\lambda C_1 = -0.8002 \;.
\end{equation}
The relative smallness of these constants (except for the last one,
which governs the amplitude $\delta z_0(0)$) may explain why the
ansatz works well down to the string.

The above saddle--point configuration is indeed situated outside of
the original contour of integration: it is a ``complex saddle point''
(although all fields have real values).  A contour deformation has to
be performed for the variables $\mi(z_1-z_1^*)(x)$, which move away
from the real axis as $|x|$ decreases.

This ansatz is tailored to the limit $L\to\infty$, but owing to the
fast decrease towards $z_{\rm vac}$, it is already quite good for
short lattices.  In Fig.~\ref{bar120-Q0m000} (bottom), we show a
logarithmic plot of the deviations of the fields from the homogeneous
vacuum, for a lattice of total length $L=20$, as well as the values
predicted by the above ansatz.

This configuration might be called a non--topological soliton,
cf.~\cite{Bochum}.  Its characteristic length (in units of the lattice
spacing $a$) is $\lambda^{-1} = 0.5673$, and its mass will be computed
in the next section.

This real scalar configuration is just one point on a
$\Gr[N_f]{U}$--manifold of solutions, obtained by the action
\eqref{chiralaction} of the chiral symmetry group.  For a generic
point on this manifold, the configuration is staggered in time.  These
solutions are saddle--point configurations for $N_f = 2$, and do not
depend on the number of colors $N_c$ (which does not appear in the
saddle--point equations).  However, the value of the action for these
configurations does depend on $N_c$; see the next section.

\subsubsection{Topological baryon, chiral limit}\label{topobaryonchiral}

We also obtained topologically nontrivial configurations,
characterized by a nonvanishing winding number $Q_w$.  In
Fig.~\ref{bar120-Q1m000} we plot a solution in the baryon sector with
$Q_w = 1$.  All fields are scalar, and have the symmetries described
in Section \ref{vacuum1+1}.

Away from the string ($x \gg 1$), the moduli of the fields are close
to $z_{\rm vac}$, and the phase varies linearly.  In this region, we
can apply the linear theory described in Section \ref{linear}, in
particular the ansatz \eqref{scalar-Ansatz-nlin}.  Now only the
coefficient $c_{+r}$ has to vanish, to prevent the deviations from
exploding as $x\to\infty$.  As in the vacuum case, the coefficient
$c_{-i}$ can take any value, yielding only a chiral shift of the
fields.  We find that $c_{+i}=\mi\alpha$ is purely imaginary, as in
the vacuum.  The fields are well fitted by
\begin{equation}\label{topobaryon}
  v_1(x+1/2) = \isq3\e^{\mi\alpha(x+1)}(1+c_{-r} \, (-\e^{-\lambda})^x)
  \quad{\rm etc.}
\end{equation}
for positive $x$. Near the string, the moduli of the fields behave in
a similar way as in the non--topological sector, whereas their phases
make a small jump $\beta$ at $x = 0$.  Both this jump and the value of
$c_{-r}$ can be computed from the full saddle--point equations near
the string (see below).  The value of the parameter $\alpha$ depends
on the height of this jump: $\alpha$ will not be exactly equal to its
value in the vacuum, which is $2\pi/L$ for this topological sector,
but rather to $(2\pi-\beta)/L$.  As a consequence, the convergence of
the fields towards the vacuum configuration away from the string will
not be exponential, but only linear (the fields coincide at the
``antipode'' of the baryon, $x = L/2$).

From Fig.~\ref{bar120-Q1m000}, we can assume that the above ansatz is
still a good approximation for $v_1(1/2)$, and for $v_1^*(1/2)$ up to
a phase jump of $\beta/2$.  Given this assumption and setting $N_f =
2$, the saddle--point equation on the baryon worldline,
$v_0(0)+1/v_0^*(0) = 1/2\{v_1(1/2) + 1/v_1^*(1/2)\}$, yields two real
equations, one of which reads
\begin{displaymath}
  \frac{\sin\big(3\alpha/2+\beta/2)}{\sin(\alpha/2)} = \frac{3(1+C_1)}
  {2(1+C_1\e^\lambda)} \;.
\end{displaymath}
For small angles $\alpha$ and $\beta$, this equation gives a linear
relation between them: using the value for $C_1$ obtained in the
non--topological sector, we get $\beta \approx 3.24\ \alpha$. The
value of the slope $\alpha$ then is $\alpha = 2\pi Q_w /(3.24 + L)$.

\subsubsection{Topological baryon, $m \not= 0$}

As in the vacuum sector, our results for a finite quark mass are
mostly numerical (see Figs.~\ref{bar120-Q1m005}, \ref{bar120-Q2m005}).
We obtained solutions of the saddle--point equations with various
winding numbers, which are close to the corresponding vacuum
configurations except in the vicinity of the baryon string.  Near the
antipode of the string ($x = L/2$), the field approaches the
homogeneous value $z_+$, and the fields can be fitted by the linear
theory developed in Section \ref{linear-m}.

\section{Mass of the static baryon ($N_f = 2$)}\label{sec:mass}

The mass $M_{\rm baryon}$ of the baryon is defined by comparing the
static baryon partition function to the vacuum one.  Since the
saddle--point configurations are classified according their winding
number $Q_w$, the comparison is performed within a given topological
class, i.e.~between a baryon configuration and the corresponding
vacuum sector.

In the limit of large lattices, the ratio of partition functions is
expected to behave as
\begin{equation}
  \frac{{\cal Z}_{\text{baryon},Q_w}(L\times T)}{{\cal Z}_{{\rm
        vacuum},Q_w}(L\times T)} \propto {\rm e}^{-M_{{\rm
        baryon},Q_w}T}\quad{\rm as}\ T\to\infty \;.
\end{equation}
As explained in Section \ref{saddlepoint}, we estimate both partition
functions through their respective lowest--order saddle--point
approximations:
\begin{equation}
\frac{\mathcal{Z}_{\text{baryon}}}{\mathcal{Z}_{\rm vacuum}} 
\sim \exp\big(-N_c \{S_{\rm baryon}[z_{\rm baryon}]
- S_{\rm vacuum}[z_{\rm vacuum}]\}\big).
\end{equation}
As before, we will only treat the $(1+1)$--dimensional case.

\subsection{Mass of the static non--topological baryon}

We first study the baryonic excitation on top of a homogeneous vacuum,
which has vanishing winding number.  The value of the action for the
homogeneous vacuum was given in Eq.~\eqref{valueS}.  It is
proportional to the volume of the lattice, and is called a ``sea''
term in the literature.

The action $S_{\rm vacuum}$ is part of the full static baryon action
\eqref{staticbaryonaction}, which leads to a ``sea'' contribution to
the baryon mass:
\begin{displaymath}
  M_{\rm sea}\defi N_c\ \frac{S_{\rm vacuum}[z_{\rm baryon}] -S_{\rm
      vacuum}[z_{\rm vac}]}{T} \;.
\end{displaymath}
We now assume the limit $m = 0$ (Section \ref{baryon-chiral}).  By the
time--independence of both configurations, the sea contribution reads
\begin{displaymath}
  M_{\rm sea} = N_f N_c \left(\sum_{x = -L/2}^{L/2} \ln
    \frac{N_1(x+1/2)N_0(x)} {M(x)}_{\big|Z=z_{\rm baryon}} - \ln
    \frac{N_1(x+1/2)N_0(x)} {M(x)}_{\big|Z=z_{\rm vac}}\right) \;.
\end{displaymath}
Since the deviations of the fields from $z_{\rm vac}$ are small and
decrease exponentially --- see Eq.~\eqref{ans-z} --- it is reasonable
to keep only the linear order, and extend the sums to $L = \infty$.
Quite remarkably, this linear approximation gives a \emph{vanishing}
sea term (for $m = 0$).  Alternatively, we can compute the sea term
from the $N_f = 2$ configuration obtained numerically.  In this way we
obtain
\begin{displaymath}
  M_{\rm sea} \approx 0.02324 \times N_c \quad \mbox{(in units of
    $a^{-1}$)} \;.
\end{displaymath}
This answer is small ($5\%$) compared to the second term we compute
below, so the linear approximation (giving $M_{\rm sea} = 0$) is
rather good in this respect.

The remaining contribution to the baryon mass comes from the sum over
traces of $\mathsf{G}$, which involve only the fields on or adjacent
to the string.  In the QCD context this is generally referred to as
the ``valence quark contribution'' \cite{Bochum}.  For the
time--independent configuration described in Section
\ref{baryon-chiral}, the valence term for $N_f = 2$ becomes
\begin{equation}\label{valenceterm}
  M_{\rm valence} = \ln \left\{\frac{\alpha_0} {\alpha_1}
    \binom{N_c+N_f-1}{N_c}\right\}\ +\ N_c\ \ln \left(\frac{2z_0(0) +
      2z_1(1/2)} {1+z_0(0)^2}\right)_{\big| z_{\rm bar}} \;.
\end{equation}
The term proportional to $N_c$ evaluates to
\begin{displaymath}
  \ln \sqrt{3}\ +\ \ln\left(\frac{1+C_1/2+\tilde C_0/2}{1+\tilde C_0/2
      +\tilde C_0^2/4}\right)\approx \ln \sqrt{3}\ +\ C_1/2 =
  0.5493\;-\;0.0485 \;.
\end{displaymath}
We notice that the second term due to the deviations $|z_{\rm
  baryon}-z_{\rm vac}|$ is small compared to the first term
$\ln\sqrt{3}$, obtained by inserting the vacuum configuration $z_{\rm
  vac}$ into \eqref{valenceterm}.

The result of this linear approximation is very close to the exact
(numerical) value, $0.5004$.  On including the combinatorial and
normalization terms (see Eq.~\eqref{alpharatio}), we finally get, for
the non--topological static baryon in the chiral limit:
\begin{align}
  M_{\rm baryon}&\approx 0.5236\, N_c +\ln(1+N_c/2) + \ln(1+N_c) \;.
\end{align}
This yields for example
\begin{align*}
  N_c = 2 &: M_{\rm baryon}=2.839\;a^{-1} \;,\\ N_c = 3&: M_{\rm
    baryon} = 3.873 \; a^{-1} \;,
\end{align*}
where we have reinstated the mass scale given by the lattice constant.

\subsection{Masses of topological baryons}

To compute the baryon masses in the topologically nontrivial sectors,
we first need to evaluate $S_{\rm vacuum}$ on the corresponding vacua
with winding number $Q_w$.  This is straightforward in the chiral
limit, where we have the accurate approximation
\eqref{result-nlin-vac} at our disposal.  The result up to second
order in the small parameter $\alpha = 2\pi Q_w / L$ is
\begin{displaymath}
  T^{-1} \left( S_{\rm vacuum}[z_{\rm vac, \alpha}] - S_{\rm vacuum}
    [z_{\rm vac}] \right) = N_f L \frac{3\alpha^2}{16} = \frac{3\pi^2
    Q_{\rm w}^2}{4 L} N_f \;.
\end{displaymath}
The vacuum energies are obtained from this by multiplication with the
number of colors, $N_c$.  They agree well with the values computed
numerically for $L = 120$ (and $N_c = 2$): $E^{(0)}_1 = 0.25$,
$E^{(0)}_2 = 0.98$.

The baryon mass in each topological sector is defined relative to the
corresponding vacuum energy $E^{(0)}_{Q_{\rm w}}$.  For a nonvanishing
quark mass, both the vacuum energy and the baryon energy are computed
numerically.  In the following table, we summarize our results for a
lattice $L = 120$, with $N_c = N_f = 2$, for various quark masses:
\begin{center}
\begin{tabular}{|r|c|c|c|}
  \hline $Q_w$ &$\quad$ $m=0$ $\quad$ & $\quad$ $am=0.002$ $\quad$&
  $\quad$ $am=0.01$ $\quad$\\ \hline 0 & 2.839 & 2.840 & 2.852 \\ 
  \hline 1 & 2.840 & 3.097 & 3.692 \\ \hline
\end{tabular}
\end{center}
In Fig.~\ref{masses}, we plot the baryon masses in the topological
sectors $Q_w = 0$ and $Q_w = 1$ as a function of the quark mass.

\section{Zigzag baryon}\label{sec:zigzag}

In the standard formulation of the theory on the Euclidean
2--dimensional square lattice, the temporal and spatial directions are
given by the lattice generators, $\hat t=(1,0)$, $\hat x =(0,1)$.
However, it is also possible to use different spacetime axes.  For
instance, on the same square lattice, we define the \emph{zigzag}
spacetime axes as $\hat t=(1/2,-1/2)$, $\hat x=(1/2,1/2)$ (see
Fig.~\ref{baryon-zigzag}).  The unit spacetime separation now has the
length $a/\sqrt{2}$.  For convenience, we choose the spacetime origin
on the \emph{middle of a link} (say, a link in the direction $\hat
t+\hat x $), so that the coordinates of lattice sites will be
half--integers, while links will be indexed by integers.

The division of the square lattice into two sublattices is now
expressed only in terms of the spatial coordinate: the links at
\emph{even} positions $x = 2n$ carry the fields $V,\ V^\dagger$, while
the links at odd $x = 2n+1$ carry the fields $W,\ W^\dagger$.

The vacuum effective action and its corresponding saddle--point
equations are still given by the formulas (\ref{vac-s.p.z},
\ref{vac-s.p.zbar}), after a suitable change of labels for links and
sites.

\subsection{Time--independent vacua}

Once again, we make a scalar time--independent ansatz for the fields:
\begin{align}
  \forall\ x\ {\rm even},\ \forall\ t : \qquad V(x,t) &=
  v(x)\mathbb{I} \;, \quad V^\dagger(x,t) = v^*(x) \mathbb{I} \;, \\ 
  \forall\ x \ {\rm odd},\ \forall\ t : \qquad W(x,t) &= w(x)
  \mathbb{I} \;, \quad W^\dagger(x,t) = w^*(x) \mathbb{I} \;.
\end{align}
In particular, this implies the equality of fields situated on links
$(x,t)$ and $(x,t+1)$.  There are 2 scalar variables at each position.
 
With these symmetries, the vacuum saddle--point equations read:
\begin{equation}\label{sp-vac-zigzag-scalar}
\begin{split}
  x\ {\rm even} : \quad 2v(x) + 2w^*(x-1) + 2am &= v(x)+1/v^*(x) \\ 
  &=w^*(x-1)+1/w(x-1) \;, \\ x\ {\rm odd} : \quad 2w(x) + 2v^*(x-1)
  +2am &= w(x)+1/w^*(x) \\ &=v^*(x-1)+1/v(x-1) \;.
\end{split}
\end{equation}
In the chiral limit, the homogeneous vacuum configurations are still
given by $v = w^* = \e^{\mi\varphi}z_{\rm vac}$ and $v^* = w =
\e^{-\mi\varphi} z_{\rm vac}$.  For fields close to this vacuum
($v(x)=\e^{\mi\varphi}z_{\rm vac}\exp\{\delta v(x)\}$ etc.)  the
linearized saddle--point equations yield the following transfer matrix
equation:
\begin{equation}
  \binom{\delta z(x)}{\delta z^*(x)} = \begin{pmatrix} -3/2 &-1/2 \\ 
    1/2 &-1/2 \end{pmatrix}\binom{\delta z(x-1)}{\delta z^*(x-1)} =
  \mathsf{T}_{zig}\binom{\delta z(x-1)}{\delta z^*(x-1)} \;.
\end{equation}
The symbol $\delta z$ stands for either $\delta v$ or $\delta w$,
depending of the parity of $x$.  In contrast with Section
\ref{linear}, we now have just one transfer matrix, which relates
deviations of $v, v^*$ to deviations of $w, w^*$ and vice versa.  This
transfer matrix $\mathsf{T}_{zig}$ is related to the matrix
$\mathsf{T}_i$ described in Section~\ref{linear}.  Indeed, it acts on
the vectors $\mathsf{I}_+,~\mathsf{I}_-$ as follows:
\begin{displaymath}
  \mathsf{T}_{zig}\mathsf{I}_+ = -\mathsf{I}_+ - \mathsf{I}_- \;,
  \quad \mathsf{T}_{zig}\mathsf{I}_- = -\mathsf{I}_- \;.
\end{displaymath}
The deviations $\delta v(0)$, $\delta v^*(0)$ are parametrized by two
\emph{complex} parameters $c_{\pm}$ as
\begin{displaymath}
  \binom{\delta v(0)}{\delta v^*(0)} = c_{+}\mathsf{I}_+ + c_{-}
  \mathsf{I}_- \;.
\end{displaymath}
The deviations will then depend on position as follows ($x$ is even):
\begin{equation}\label{linear-zig}
  \binom{\delta v(x)}{\delta v^*(x)} = \binom{c_- + (x+1) c_+}{-c_-
    -(x-1)c_+} \;, \qquad \binom{\delta w(x+1)}{\delta w^*(x+1)} =
  \binom{-c_- -(x+2) c_+}{c_- + x c_+} \;.
\end{equation}

The rest of the discussion is identical to the one following
Eq.~\eqref{scalar-Ansatz-nlin}.  The coefficient $c_-$ plays the role
of a global shift, or ``generalized chiral rotation''.  If $\re c_+
\neq 0$, the absolute values of the fields vary linearly with $x$,
which is incompatible with their periodicity.  On the other hand,
taking $c_+ = \mi\alpha$ will linearly rotate the phases of the
fields, keeping them close to some vacuum configuration, as in
Eqs.~\eqref{result-nlin-vac}.  Taking for $\alpha$ a multiple of
$2\pi/L$, we obtain a topologically nontrivial configuration.

The case of broken chiral symmetry ($m\neq 0$) can be treated along
the same lines as in Section~\ref{linear-m}; the above linear
evolution in position is then replaced by an exponential one, at least
for fields in the vicinity of the value $z_+$.  The transfer matrix
takes the form $\begin{pmatrix} -z_+^{-2}/2 &-1/2 \\ 1/2 &-3z_+^2/2
\end{pmatrix}$, and has the eigenvalues $-\exp(\pm\gamma_{zig})$
associated to the eigenvectors $\binom{1}{-\exp(\pm\kappa_{zig})}$,
with the expansions
\begin{align}\label{gammazig}
  \gamma_{zig} &= \frac{\sqrt{2}}{3^{1/4}} (am)^{1/2}+{\cal
    O}((am)^{3/2}) \;, \\ \kappa_{zig} &= -\frac{2\sqrt{2}}{3^{1/4}}
  (am)^{1/2}+{\cal O}((am)^{3/2}) \;.
\end{align}
Up to this order, the exponent $\gamma_{zig}$ differs from the
corresponding exponent $\gamma$ of Eq.~\eqref{gamma-m} by a factor of
$\sqrt{2}$.  This factor actually compensates for the ratio of unit
lengths between the two frameworks ($a$ versus $a/\sqrt{2}$).  For the
quark mass $am = 0.005$ used in compiling the figures, we have
$\gamma_{zig} = 0.0762$ and $\e^{\kappa_{zig}} = 0.85905$.

\subsection{Zigzag baryon}

In the new labeling conventions, the worldline of a static baryon
situated at position $x = 0$ forms a zigzag curve (see
Fig.~\ref{baryon-zigzag}).  Assuming that the fields are scalar and
time--independent, the $\mG$ matrix appearing in the baryonic part of
the action is
\begin{displaymath}
\begin{split}
  \mG = \{&(2 v(0) + 2 w^*(-1)+2am)^{-1}(1 + v(0)v^*(0))\nonumber\\
  &(2 w(1) + 2 v^*(0)+2am)^{-1}(1+v(0)v^*(0))\}^{T/2}\ \mathbb{I}.
\end{split}
\end{displaymath}
The resulting saddle--point equations on the string read 
\begin{equation}\label{sp-bar-zigzag-scalar}
\begin{split}
  2v(0) + 2w^*(-1) + 2am &= v(0)+1/v^*(0) \\ 
  &= (1-N_f^{-1}) \{w^*(-1) + 1/w(-1)\} \;, \\ 
    2w(1) + 2v^*(0) + 2am &= v^*(0)+1/v(0) \\ 
    &= (1-N_f^{-1}) \{w(1) + 1/w^*(1)\} \;.
\end{split}
\end{equation}
The saddle--point equations are invariant under the transformation
\begin{equation}\label{mirror}
  w(-x) \longleftrightarrow w^*(x) \;, \qquad v(-x)
  \longleftrightarrow v^*(x) \;,
\end{equation}
which is also a symmetry of our numerical solutions.

In the chiral limit, the linear dependence in \eqref{linear-zig} makes
it impossible for the absolute values of the fields to approach $z_{
  \rm vac}$ at infinity, unless the deviations are purely imaginary;
this latter possibility ($|z(x)|\equiv z_{\rm vac}$) is incompatible
with the saddle--point equations on the baryon string
\eqref{sp-bar-zigzag-scalar}.  However, for a finite lattice, infinity
is the ``antipodal point'' $x_{\infty} = L/2$.  In numerical searches
(see Fig.~\ref{zig40-Q0}, top), we found a solution which comes close
to $z_{\rm vac}$ near the antipode (but does not converge
exponentially to it).  For $x$ even, $0 < x < L/2$, the fields are
well described by
\begin{equation}
\begin{split}
  v(x) &= z_{\rm vac}\e^{c_+(x+1-L/2)} \;, \\ v^*(x) &= z_{\rm vac}
  \e^{ -c_+(x-1-L/2)} \;, \\ w(x+1) &= z_{\rm vac}\e^{-c_+(x+2-L/2)}
  \;, \\ w^*(x+1) &= z_{\rm vac}\e^{c_+(x-L/2)} \;,
\end{split}
\end{equation}
where the value of the real coefficient $c_+$ is small (for a lattice
of length $L = 80$, we found $c_+ \approx 0.04$).  The fields in the
region $-L/2 < x < 0$ are obtained by applying the symmetry
\eqref{mirror}.  At all points $x \not= 0$, the fields are close to a
``generalized homogeneous vacuum''.  The coefficient $c_+$ is fixed by
using the above ansatz up to $x = -1$, and then enforcing the
equations \eqref{sp-bar-zigzag-scalar}: in this way one obtains (to
lowest order in $c_+$) the transcendental equation
\begin{equation}\label{implicitc+}
  c_+ \e^{c_+ L} = 1 \;,
\end{equation}
with approximate solution $c_+ \approx L^{-1} \ln L$.  The field on
the baryon takes the value
\begin{equation}\label{v(0)}
  v(0) = v^*(0) = \frac{3z_{\rm vac}}{2} \sqrt{c_+} \;.
\end{equation}
This configuration is very ``distorted'' compared to the original
contour of integration.  Indeed, the ratios $v / v^*$ and $w / w^*$
are of order $L$ in the vicinity of the string.  In the case $L = 80$,
the above equations yield $c_+ = 0.04018$ and $v(0) = 0.1736$, in
excellent agreement with our numerical data.

\subsubsection{Zigzag baryon, broken chiral symmetry}\label{zigbroken}

If chiral symmetry is broken by a nonvanishing quark mass, the
deviations from $z_+$ evolve exponentially with $x' = x + L/2$ away from
the antipodal point $x' = 0$.  More precisely, we should have, in some
domain $|x'| \ll L/2$,
\begin{equation}\label{ansatzzig}
  \binom{\delta z(x')}{\delta z^*(x')} \approx \binom{\eps_+
    (-\e^{\gamma_{zig}})^{x'} + \eps_- (-\e^{-\gamma_{zig}})^{x'}} {-
    \eps_+\, e^{\kappa_{zig}}(-\e^{\gamma_{zig}})^{x'} - \eps_-\,
    e^{-\kappa_{zig}}(-\e^{-\gamma_{zig}})^{x'}} \;.
\end{equation}
From the mirror symmetry \eqref{mirror}, the coefficients $\eps_{\pm}$
are related by
\begin{equation}\label{mirror2}
  \eps_- = -\e^{\kappa_{zig}}\,\eps_+ \;.
\end{equation}
We numerically computed a solution with $L = 80$, $am = 0.005$ (see
Fig.~\ref{zig40-Q0}, bottom), for which this ansatz works well up to
the string.  Using this fact, it is possible to estimate the value of
$\eps_+$ (the last remaining parameter), as in the chiral case.  The
crudest approximation yields the equation
\begin{displaymath}
  -\eps_+ \,\kappa_{zig} C\,\e^{2C\eps_+} = 2 \;, \quad \mbox{with}
  \quad C\defi \exp\{\gamma_{zig}(L/2-1)\} \;.
\end{displaymath}
For our configuration, the solution of this equation is $\eps_+ =
0.0614$.  This configuration has a mass $M = 2.885\ a^{-1}$ with
respect to the homogeneous vacuum.

There also exist topologically nontrivial vacuum and baryon
saddle--point configurations with the symmetry \eqref{mirror}.  For
example, Fig.~\ref{zig80log} shows the solution in the sector $Q_w =
1$ with quark mass $am = 0.005$ on a lattice of length $L = 160$.
With respect to the corresponding vacuum configuration, this
configuration has a mass $M = 3.18\ a^{-1}$.

\section{Concluding remarks}

The ``color--flavor transformation'' introduced in
\cite{circular,icmp97} replaces an integral over the gauge group ${\rm
  U}(N_c)$ by an integral over the ``flavor'' degrees of freedom.  In
the present paper we extended this transformation to the gauge group
${\rm SU} (N_c)$.

The color--flavor transformation can be interpreted as a kind of
duality, linking two different formulations of the theory.  We believe
that this duality transformation may be useful for treating realistic
non--perturbative QCD.  Here we have applied it to a simple model of
two--dimensional lattice fermions.  The non--Abelian theory we have
treated is of course too far from realistic four--dimensional QCD for
our results to be of direct phenomenological relevance.

The main approximation we made was to assume the strong--coupling
limit for the lattice gauge fields.  In this approximation gluons do
not propagate, and the connection to asymptotic freedom at short
distances is lost.  An unphysical consequence is the absence of the
${\rm U}(1)$ chiral anomaly.  Thus, the chiral symmetry group of our
low-energy effective action is not ${\rm SU}(N_f)$ but the larger
group ${\rm U}(N_f)$.

Among the gauge groups ${\rm SU}(N_c)$ the case $N_c = 2$ is special,
as the vector and covector representations of ${\rm SU}(2)$ happen to
be equivalent.  Since these representations correspond to quarks and
antiquarks respectively, there is no physical distinction between
baryons and mesons in that case.  This symmetry between baryons and
mesons is obscured in the present treatment which, by the use of a
saddle--point approximation valid only for $N_c \gg 1$, is geared to
the large--$N_c$ limit.  It can, however, be made manifest by
identifying ${\rm SU}(2)$ with the compact symplectic gauge group
${\rm Sp}(2)$ and using the color--flavor transformation for the
latter \cite{icmp97}.

To extend the formalism to lattice QCD in four dimensions, we need to
take into account the spin degrees of freedom and put the chiral
fermions properly on the lattice.  We hope to address these issues in
a separate publication.  Here we only note that a first step towards a
more realistic color--flavor transformed theory of the strong
interaction was described in \cite{BS,BS-1}, where we discuss the
effect of spontaneous chiral symmetry breaking and estimate the
numerical values of the chiral condensate, the pion decay constant and
the mass of the pion.

\medskip\noindent{\bf Acknowledgements}.

We are grateful to A.~Altland, A.~Andrianov, D.~Diakonov, A.~Morel,
V.~Petrov, P.~Pobylitsa, M.~Polyakov, R.~Seiler, A.~Smilga and A.~Wipf
for useful discussions and comments.

\appendix
\section{Action of the color and flavor groups}
\label{appA} 

Transformations $U \in \Gr[N_c]{GL}$ of color space act on the
one--fermion operators by
\begin{alignat}{2}
  T_U \f^k_{+a} T_U^{-1} &= \f^j_{+a} {(U^{-1})}^{kj} \;, \quad &T_U
  \f^k_{-a} T_U^{-1} &= \f^j_{-a} U^{jk} \;, \nonumber \\ T_U
  \fb^k_{+a} T_U^{-1} &= \fb^j_{+a} U^{jk} \;, \quad &T_U \fb^k_{-a}
  T_U^{-1} &= \fb^j_{-a} {(U^{-1})}^{kj} \;. \nonumber
\end{alignat} 
Transformations $\begin{pmatrix} A & B \\ C & D \end{pmatrix} \in
\Gr[2N_f]{GL}$ of flavor space can be decomposed in the way shown in
Eq.~(\ref{decomposition}), and the action of the various factors may
be described separately.  An element $\zeta = \begin{pmatrix} 1 & Z \\ 
  0 & 1 \end{pmatrix} = \exp\begin{pmatrix} 0 & Z \\ 0 & 0
\end{pmatrix}$ acts by
\begin{alignat}{2}
  T_\zeta \f^k_{+b} T_\zeta^{-1} &= \f^k_{+b} - \fb^k_{-a} Z_{ba} \;,
  \quad &T_\zeta \f^k_{-b} T_\zeta^{-1} &= \f^k_{-b} + \fb^k_{+a}
  Z_{ab}\;, \nonumber \\ T_\zeta \fb^k_{+b} T_\zeta^{-1} &= \fb^k_{+b}
  \;,\quad &T_\zeta \fb^k_{-b} T_\zeta^{-1} &= \fb^k_{-b}\;, \nonumber
\end{alignat}
an element $\diag(A,D) = \begin{pmatrix} A & 0 \\ 0 & D \end{pmatrix}
\in \Gr[2N_f]{GL}$ by
\begin{alignat}{2}
  T_{\diag(A,D)} \f^k_{+b} T_{\diag(A,D)}^{-1} &= \f^k_{+a}
  {(A^{-1})}_{ba} \;, \quad &T_{\diag(A,D)} \f^k_{-b}
  T_{\diag(A,D)}^{-1} &= \f^k_{-a} D_{ab} \;, \nonumber \\ 
  T_{\diag(A,D)} \fb^k_{+b} T_{\diag(A,D)}^{-1} &= \fb^k_{+a} A_{ab}
  \;, \quad &T_{\diag(A,D)} \fb^k_{-b} T_{\diag(A,D)}^{-1} &=
  \fb^k_{-a} {(D^{-1})}_{ba} \;, \nonumber
\end{alignat}
and an element $\tilde{\zeta} = \begin{pmatrix} 1 & 0 \\ \tilde{Z} & 1
\end{pmatrix} \in \Gr[2N_f]{GL}$ by
\begin{alignat}{2}
  T_{\tilde{\zeta}} \f^k_{+b} T_{\tilde{\zeta}}^{-1} &= \f^k_{+b} \;,
  \quad &T_{\tilde{\zeta}} \f^k_{-b} T_{\tilde{\zeta}}^{-1} &=
  \f^k_{-b} \;, \nonumber \\ T_{\tilde{\zeta}} \fb^k_{+b}
  T_{\tilde{\zeta}}^{-1} &= \fb^k_{+b} + \f_{-a}^k \tilde{Z}_{ab} \;,
  \quad &T_{\tilde{\zeta}} \fb^k_{-b} T_{\tilde{\zeta}}^{-1} &=
  \fb^k_{-b} - \f_{+a}^k \tilde{Z}_{ba} \;. \nonumber
\end{alignat}
All these formulas are particular cases of the fermionic Fock--space
representation of the Lie group $\Gr[2N_f]{GL}$ expounded in Chapter 9
of \cite{Perelomov}.

\section{Normalization constants}
\label{normalisation}

We are going to calculate the normalization constants $\alpha_Q^{-1} =
\int_{G} dg \, |\langle B_Q | T_g | B_Q \rangle| ^2$ introduced in
Eq.~(\ref{alphaQ}) -- for the values $Q = 1, 0, -1$.  To that end, we
employ the decomposition of the group $G = {\rm U}(2N_f)$ given by
Eqs.~(\ref{decomposition}) and (\ref{ADdecompo}).  This yields
\begin{displaymath}
  |\langle B_1 | T_g | B_1 \rangle|^2 = \frac{|(\sqrt{1 + ZZ^\dagger
      })_{1a}\mathcal{U}_{a1}|^{2N_c}} {\det(1 + ZZ^\dagger )^{N_c}}
\end{displaymath}
for $Q = 1$, and similar expressions for the other two cases.  The
first step now is to do the integral over $\mathcal{U}\in\Gr[N_f]
{U}$, which for $Q = \pm 1$ is effectively an integral over a
$(2N_f-1)$--dimensional sphere.  Carrying it out by the method of
Section \ref{integralU}, we get the preliminary expressions
\begin{align}
  \alpha_0^{-1} &= C_{N_f}\int_{\mathbb{C}^{N_f \times N_f}} \frac{dZ
    dZ^\dagger}{\det(1+Z Z^\dagger)^{2N_f+N_c}} \;, \label{firststep1}
  \\ \alpha_1^{-1} = \alpha_{-1}^{-1} &= C_{N_f}\frac{(N_f-1)!
    N_c!}{(N_c+N_f-1)!}\ \int_{\mathbb{C}^{N_f \times N_f}}
  \frac{[(1+Z^\dagger Z)_{11}]^{N_c} \; dZ dZ^\dagger} {\det(1+Z
    Z^\dagger)^{2N_f+N_c}} \;, \label{firststep2} \\ \intertext{where
    $C_{N_f}$ is defined by} C_{N_f}^{-1} &= \int_{\mathbb{C}^{N_f
      \times N_f}} \frac{dZ dZ^\dagger}{\det(1+Z Z^\dagger)^{2N_f}}
  \;. \label{CNfdef}
\end{align}
For later convenience, we have made a change of integration variables
$Z \leftrightarrow Z^\dagger$ in the numerator of the integral in
(\ref{firststep2}).

In the second step we perform the integration over the $N_f \times
N_f$ matrix $Z$ using a recursion procedure similar to that in
\cite{Hua}.  From here on we use the simplified notation $n = N_f$.
The recursion consists in slicing the matrix $Z$ into vertical
vectors, step by step.  We now detail the first step of the recursion.
We decompose $Z$ as $Z = (Z_{n,n-1},z_1)$, where $z_1$ is a (column)
$n$--vector, and $Z_{n,n-1}$ is a $n \times (n-1)$ matrix.  We then
have the expressions
\begin{eqnarray*}
  ZZ^\dagger &=& Z_{n,n-1}Z_{n,n-1}^\dagger + z_1 z_1^\dagger \;, \\ 
  \quad(Z^\dagger Z)_{11} &=& (Z_{n,n-1}^\dagger Z_{n,n-1})_{11} \;.
\end{eqnarray*}
Using the (positive definite) $n\times n$ matrix $\Gamma_1$ which is 
defined as the square root of
\begin{displaymath}
  \Gamma_1^2 = 1 + Z_{n,n-1}Z_{n,n-1}^\dagger \;,
\end{displaymath}
we make a change of variables, from $z_1$ to $w_1 = \Gamma_1^{-1}
z_1$.  From $1 + ZZ^\dagger = \Gamma_1(1 + w_1 w_1^\dagger) \Gamma_1$,
we get the relation
\begin{displaymath}
  \det(1 + Z Z^\dagger) = (1 + w_1^\dagger w_1) \det(1 + Z_{n,n-1}
  Z_{n,n-1}^\dagger) \;.
\end{displaymath}
The change of variables from $Z$ to $\{Z_{n,n-1},w_1\}$ has the
Jacobian $\det(1+Z_{n,n-1} Z_{n,n-1}^\dagger)$.  Each of the integrals
(\ref{firststep1}), (\ref{firststep2}), and (\ref{CNfdef}) can now be
written as the product of a $Z_{n,n-1}$--integral times a
$w_1$--integral.

The former can in turn be expressed as the product of a $Z_{n,
  n-2}$--integral times a $w_2$--integral (with $w_2$ a $n$--vector),
which can be decomposed in turn, and so on, until we reach, at the
$n$--th step, a $Z_{n,1}$--integral, i.e.~an integral over the first
column of the original matrix $Z$.  We call this column vector $w_n$
for reasons of homogeneity.

The successive Jacobians multiply to give the following integration
measure:
\begin{displaymath}
  dZ dZ^\dagger = dw_1^\dagger dw_1\, (1+w_2^\dagger w_2)
  dw_2^\dagger dw_2 \cdots (1+w_n^\dagger w_n)^{n-1} dw_{n}
  dw_{n}^\dagger \;.
\end{displaymath}
The integrands in (\ref{firststep1}-\ref{CNfdef}) also have simple
expressions in the new variables, due to the identities $(Z^\dagger
Z)_{11} = w_n^\dagger w_n$ and
\begin{displaymath}
  \det(1+ZZ^\dagger) = (1+w_1^\dagger w_1)(1+w_2^\dagger w_2) \cdots
  (1+w_n^\dagger w_n) \;.
\end{displaymath}
The $w_i$--integrals to be performed are all of the type ($N\geq n$)
\begin{displaymath}
  \int_{\mathbb{C}^{n}} \frac{dw^\dagger dw }{(1 + w^\dagger w)^{N+1}}
  = \pi^{n} \frac{(N-n)!}{N!} \;.
\end{displaymath}
The resulting expressions for the normalization constants are
\begin{align}
  \alpha_0 &= \frac{1}{C_{N_f}\pi^{N_f^2}} \; \frac{(2N_f+N_c-1)!
    \cdots (N_f+N_c)!} {(N_c+N_f-1)! \cdots N_c!} \;, \label{alpha0}
  \\ \alpha_1 = \alpha_{-1} &= \frac{1}{C_{N_f}\pi^{N_f^2}}\;
  \frac{N_f (2N_f+N_c-1)! \cdots (N_f+N_c+1)!} {(N_c+N_f-2)! \cdots
    N_c!} \;, \label{alpha1} \\ C_{N_f} &= \frac{1}{\pi^{N_f^2}} \;
  \frac{(2N_f-1)!  \cdots N_f!} {(N_f-1)!  \cdots 0!} \;, \label{CNf}
\end{align}
where we have reinstated $n = N_f$.  The quantity entering into the
baryon mass is the ratio
\begin{equation}\label{alpharatio}
  \frac{\alpha_1}{\alpha_0} = \frac{N_f}{N_f + N_c} \;.
\end{equation}

\section{Static baryon}
\label{staticbaryon}

In this appendix we prove the formula (\ref{staticbaryonaction}) for
the action functional of the static baryon sector.  We need to
integrate polynomials in the quark fields along the world line of the
baryon (the baryon ``string''), weighted by the same Gaussian as in
the vacuum sector.

We start out using the short--hand notation $t\hat 0 \equiv 0 + t\hat
0$ of Section \ref{sec:statbar} for sites and links on the string.
The part of the integrand containing the quark fields situated on the
string, namely $\psi(t \hat 0)$, $\psib(t\hat 0)$ for $t = 0, \ldots,
T-1$, then reads
\begin{multline}
  \chi_1\big(\psib(0), N({\Ss \frac{1}{2}\hat 0}) \psi(1 \hat 0)\big)
  \; \chi_1\big(\psib(1 \hat 0), N({\Ss \frac{3}{2}\hat 0}) \psi(2\hat
  0)\big) \cdots \\ \cdots \chi_1\big(\psib({\Ss (T-1)\hat 0}), N({\Ss
    (T-\frac{1}{2})\hat 0})\psi(T\hat 0)\big) \times \exp - \sum_{t =
    0}^{ T-1} \psib^i(t \hat 0) M(t\hat 0)\psi^i(t\hat 0) \;.
  \nonumber
\end{multline}
Isolating the terms with fermions at the site $n = t\hat 0$, we are
faced with the integral
\begin{align*}
  &\int d\psib(n) d\psi(n) \ \chi_1\big(\psib(n - \hat0), N({\Ss n -
    \hat 0/2})\psi(n)\big) \; \e^{- \psib(n) M(n) \psi(n)} \;
  \chi_1\big( \psib(n),N({\Ss n+\hat 0/2})\psi(n+\hat 0)\big) \\ 
  = &\left( \alpha_1 \frac{(N_f-1)!}{(N_c+N_f-1)!} \right)^2 \int
  \prod_{i,a} d\psib^i_a(n) d\psi^i_a(n) \, \e^{- \psib_c^k(n) 
    M_{cc'}(n) \psi_{c'}^k(n)} \; \\ 
  &\times\sum_{\sigma, \tau \in \mathfrak{S}_{N_c}} \sgn\sigma
  \sgn\tau \, \prod_i\psib^i_a(n-\hat 0)N_{ab}({\Ss n-\hat 0/2})
  \psi^{\sigma(i)}_b(n) \prod_j\psib^j_{a'}(n)N_{a'b'} ({\Ss n + \hat
    0/2})\psi^{\tau(j)}_{b'}(n+\hat 0) \\ 
  = &\left( \alpha_1 \frac{(N_f-1)!}{(N_c+N_f-1)!} \right)^2
  \sum_{\sigma, \tau \in \mathfrak{S}_{N_c}} \sgn(\sigma\tau) \,
  \prod_i \Big[ \psib^{\sigma^{-1}(i)}_{a_i}(n-\hat 0)N_{a_i b_i}
  ({\Ss n - \hat 0/2}) \\ 
  &\Big\{ \int d\psib^i(n) d\psi^i(n) \ \psi^i_{b_i}(n)
  \psib^i_{a'_i}(n) \e^{- \psib^i(n)M(n)\psi^i(n)}\Big\} N_{a'_i b'_i}
  ({\Ss n+\hat 0/2})\psi^{\tau(i)}_{b'_i}(n+\hat 0) \Big] \;,
\end{align*}
where the first equality sign uses the expression (\ref{chi1}) for the
function $\chi_1$.  Note that the integral between curly brackets
involves only fermions of color $i$.  The fermionic version of Wick's
theorem yields for it the value $M^{-1}_{b_i a'_i}(n) \det M(n)$, so
after combining the permutations $\sigma$ and $\tau$, the above
expression becomes
\begin{align*}
  &\alpha_1^2 \frac{(N_f-1)!^2}{(N_c+N_f-1)!^2} \; N_c! \det
  M(n)^{N_c} \sum_{\rho\in \mathfrak{S}_{N_c}} \sgn\rho \, \prod_i
  \psib^i_{a_i} (n - \hat 0) G_{a_i b'_i} ({\Ss n - \hat 0 \to n+ \hat
    0}) \psi^{\rho(i)}_{b'_i} (n + \hat 0) \\ &= \alpha_1 \det
  M(n)^{N_c} \, \binom{N_c + N_f-1}{N_f - 1}^{-1} \chi_1 \big( \psib(n
  - \hat 0), G({\Ss n -\hat 0 \to n +\hat 0}) \psi(n+\hat 0) \big) \;,
\end{align*}
with the ``propagator'' $G({\Ss n-\hat 0\to n+\hat 0}) \defi N({\Ss n
  - \hat 0/2}) M(n)^{-1} N({\Ss n + \hat 0/2})$.

Repeating the procedure, we successively integrate over the quark
fields along the string, by which process the matrices $N$ and $M$ get
organized into a single propagator.  In the final integration step, we
need to take into account the periodic boundary conditions for the
quark fields: $\psi(T \hat 0) = \psi(0)$.  The final integral over
$\psi(0)$ then reads
\begin{displaymath}
  \int d\psib(0) d\psi(0) \, \chi_1\big(\psib(0), G({\Ss 0\to T\hat
    0}) \psi(0)\big) \; \e^{- \psib(0) M(0) \psi(0)} \;.
\end{displaymath}
We now use the following expression for the function $\chi_1$:
\begin{displaymath}
  \chi_1(\phib,\phi) = \alpha_1 \frac{(N_f - 1)!} {(N_c +
    N_f-1)!} \sum_{\{a_i\}} \sum_{\sigma \in \mathfrak{S}_{N_c}}
  \prod_{i=1}^{N_c} \phib^i_{a_i} \phi^i_{a_{\sigma(i)}} \;,
\end{displaymath}
which is easily obtained from Eq.~(\ref{chi1}) by interchanging the
product over colors with the sum over flavors.  Wick's theorem 
then yields for the $\psi(0)$--integral the result
\begin{displaymath}
  \alpha_1 \frac{(N_f-1)!} {(N_c+N_f-1)!} \det M(0)^{N_c} (-1)^{N_c}
  \sum_{ \{a_i, b_i\} } \sum_{\sigma\in \mathfrak{S}_{ N_c}} \prod_i
  G_{a_{\sigma (i)} b_i}({\Ss 0\to T\hat 0}) M^{-1}_{b_i a_i}(0) \;.
\end{displaymath}
The last matrix product may also be expressed in terms of the
propagator $\mathsf G$ defined in Eq.~(\ref{propagatorworldline}),
$\mathsf G = G({\Ss 0\to T\hat 0}) M(0)^{-1}$.

What's the interpretation of the sign factor $(-1)^{N_c}$?  To answer
that question, recall that we evaluated the Grassmann field integral
using time--\emph{periodic} boundary conditions (instead of the
conventional time-antiperiodic ones).  In a $d$--dimensional quantum
mechanical frame work with Hamiltonian $H$ and inverse temperature
$\beta$, this would mean that we are computing not the usual partition
function but rather the \emph{super}trace ${\rm Tr} \, (-1)^{N_{\rm
    F}} {\rm e}^{- \beta H}$ with $N_{\rm F}$ the total fermion
number.  The overall sign factor $(-1)^{N_c}$ originates from that
very fermion number, and is simply telling us that the baryon is a
fermion (boson) if $N_c$ is odd (resp.~even).

Let us take a closer look at the contributions from the sum over
permutations $\sigma\in \mathfrak{S}_{N_c}$.  Each permutation
$\sigma$ can be uniquely decomposed into a product of independent
cycles.  Denoting by $c_l(\sigma)$ the number of cycles of length $l$
in this decomposition, the contribution from $\sigma$ to the partition
function can be written as
\begin{displaymath}
  \sum_{\{ a_i \}} \prod_{i=1}^{N_c} \mathsf{G}_{a_i a_{\sigma(i)}} =
  \prod_{l=1}^{N_c} (\Tr \mathsf{G}^l)^{c_l(\sigma)}.
\end{displaymath}
The permutation group $\mathfrak{S}_{N_c}$ may be partitioned into
disjoint classes with respect to conjugation ($\sigma, \sigma'$ are
said to be conjugate to each other iff there exists a permutation
$\tau$ such that $\sigma' = \tau^{-1} \sigma \tau$).  Two permutations
$\sigma$ and $\sigma'$ are in the same conjugacy class iff they have
the same cycle structure, i.e.~$\forall l: \, c_l(\sigma) =
c_l(\sigma')$.  This allows to rewrite the sum over $\sigma$ as a sum
over the conjugacy classes $\hat{\sigma} \in \hat{\mathfrak S}_{N_c}$,
taking into account the cardinality of each class, $\mathcal{N}(\hat
\sigma)$, given in Eq.~(\ref{normalizationconstants}).  We then obtain
the result (\ref{staticbaryonaction}).

\newpage\thispagestyle{empty}
\begin{figure}[htb]\lbfig{cf-1}
\begin{center}
  \setlength{\unitlength}{1cm}
\begin{picture}(10,12.5)
  \put(-1.7,-0.0) {\mbox{\epsfysize=10.0cm\epsffile{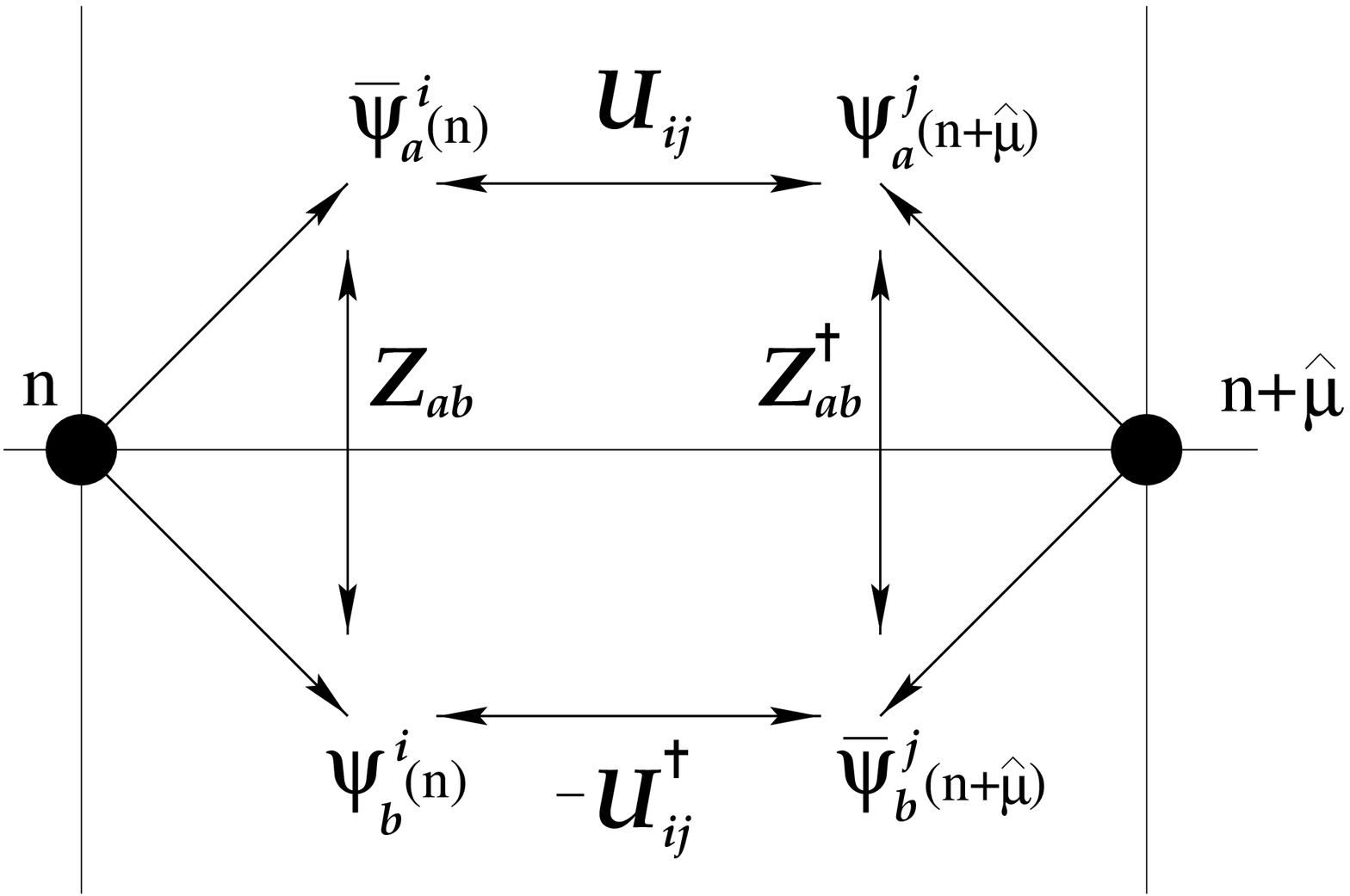}}}
\end{picture}
\caption{Coupling of the fermion fields before and after the 
  color--flavor transformation.}
\end{center}
\end{figure}
\newpage\thispagestyle{empty}
\begin{figure}[htb]
\begin{center}
  \setlength{\unitlength}{1cm}
\begin{picture}(10,12.5)
  \put(-3,-0.0) {\mbox{\epsfysize=12.0cm\epsffile{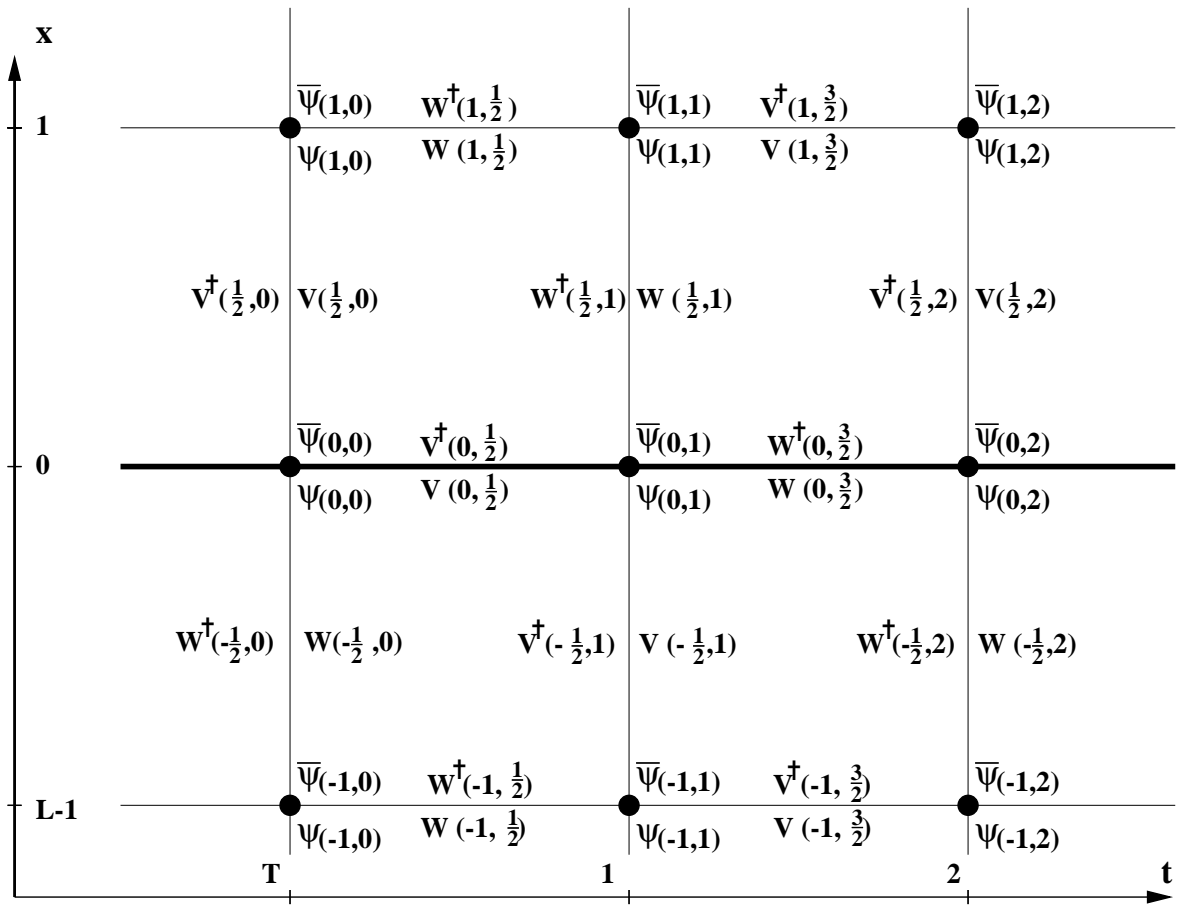}}}
  \lbfig{grid-bar}
\end{picture}
\caption{Baryon string placed on a two--dimensional lattice.}
\end{center}
\end{figure}
\newpage\thispagestyle{empty}
\begin{figure}[htb]
\begin{center}
  \setlength{\unitlength}{1cm}
\begin{picture}(11,21)
  \put(-2,0) {\mbox{\epsfysize=24.0cm\epsffile{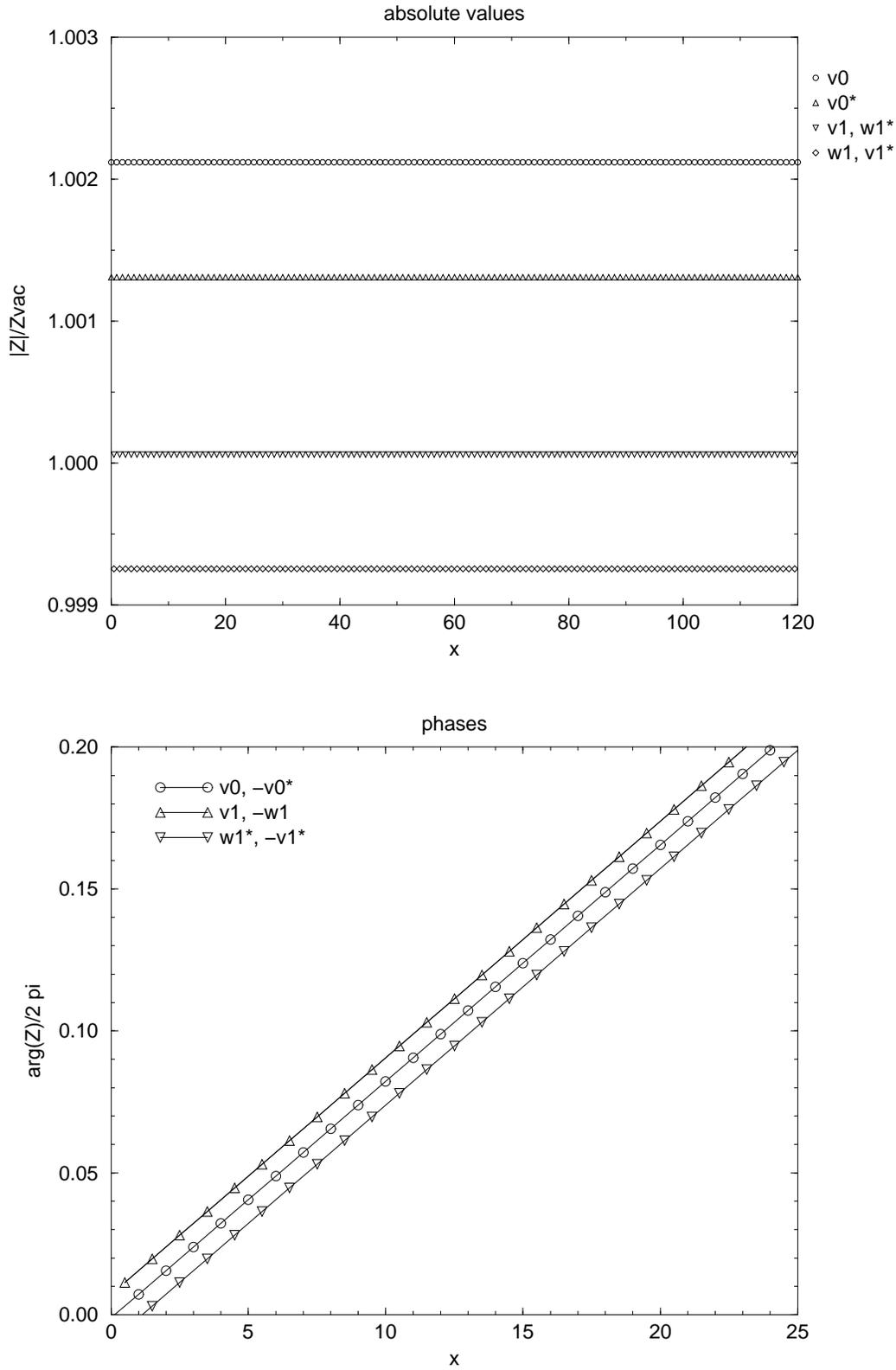}}}
  \lbfig{vac120-Q1m000}
\end{picture}
\caption{Vacuum configuration computed numerically, with winding number 
  $Q_w = 1$ in the chiral limit.  Fields which are numerically
  indistinguishable (e.g.~$|v_1|$ and $|w_1^*|$) are represented by
  the same symbol.  There is a perfect fit with formulas
  \eqref{result-nlin-vac}, including the $\alpha^2$ correction.  The
  difference between $|v_0|$ and $|v_0^*|$ comes from $\re c_{-i}\neq
  0$.}
\end{center}
\end{figure}
\newpage\thispagestyle{empty}
\begin{figure}[htb]
\begin{center}
  \setlength{\unitlength}{1cm}
\begin{picture}(11,21)
  \put(-2.0,0.5) {\mbox{\epsfysize=23.0cm\epsffile{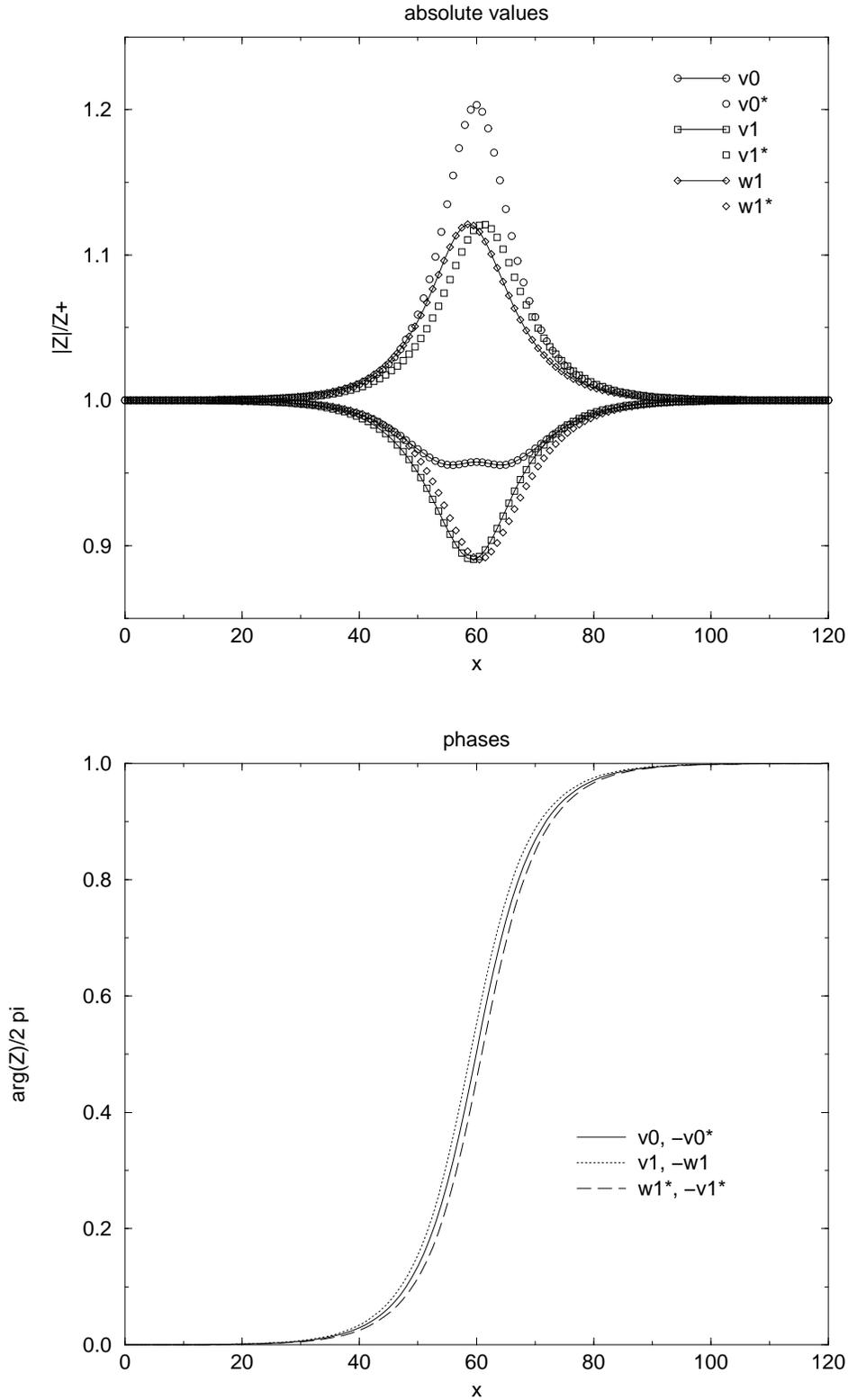}}}
  \lbfig{vac120-Q1m005}
\end{picture}
\caption{Numerical vacuum configuration with winding number 
  $Q_w = 1$ and chiral symmetry explicitly broken by a finite quark
  mass.  The fields are normalized with respect to the corresponding
  value $z_+$.  The phases $\arg(v_0)$ and $-\arg(v_0^*)$ are
  indistinguishable, so we plot them together (idem for $\arg(v_1)$
  and $-\arg(w_1)$, resp.~$\arg(w_1^*)$ and $-\arg(v_1^*)$).}
\end{center}
\end{figure}
\newpage\thispagestyle{empty}
\begin{figure}[t]
\begin{center}
  \setlength{\unitlength}{1cm}
\begin{picture}(11,21)
  \put(-2.0,0.5)
  {\mbox{\epsfysize=16.0cm\epsffile{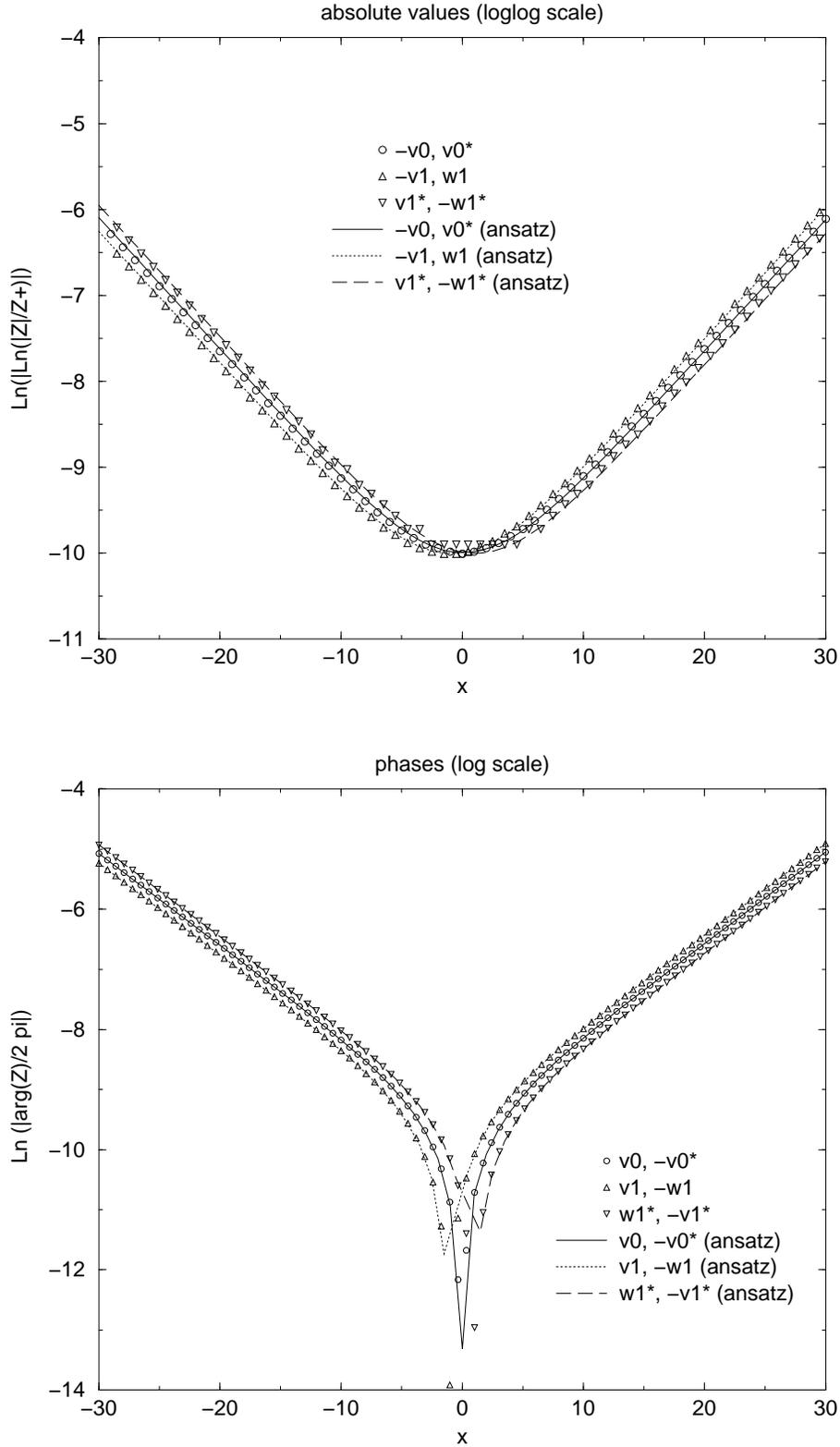}}}
  \lbfig{vac120-Q1m005log}
\end{picture}
\caption{Same vacuum configuration as in the previous figure, plotted on
  a logarithmic scale.  We fit the fields in the range $0 < |x| < 30$
  (where they are close to $z_+$) with the ansatz
  \eqref{result-m-vac}, using the theoretical values for $\gamma$,
  $\kappa_1$, $\kappa_{\pm 0}$ from Eqs.~\eqref{datam005}.  The best
  fit is obtained with $\eps_+ = (-2.80 + \mi 7.62)\times 10^{-5}$ and
  $\eps_- = (-1.82 - \mi 5.48)\times 10^{-5}$.}
\end{center}
\end{figure}
\newpage\pagestyle{empty}
\begin{figure}[htb]
\begin{center}
  \setlength{\unitlength}{1cm}
\begin{picture}(10,21)
  \put(-2.5,0.0)
  {\mbox{\epsfysize=17.0cm\epsffile{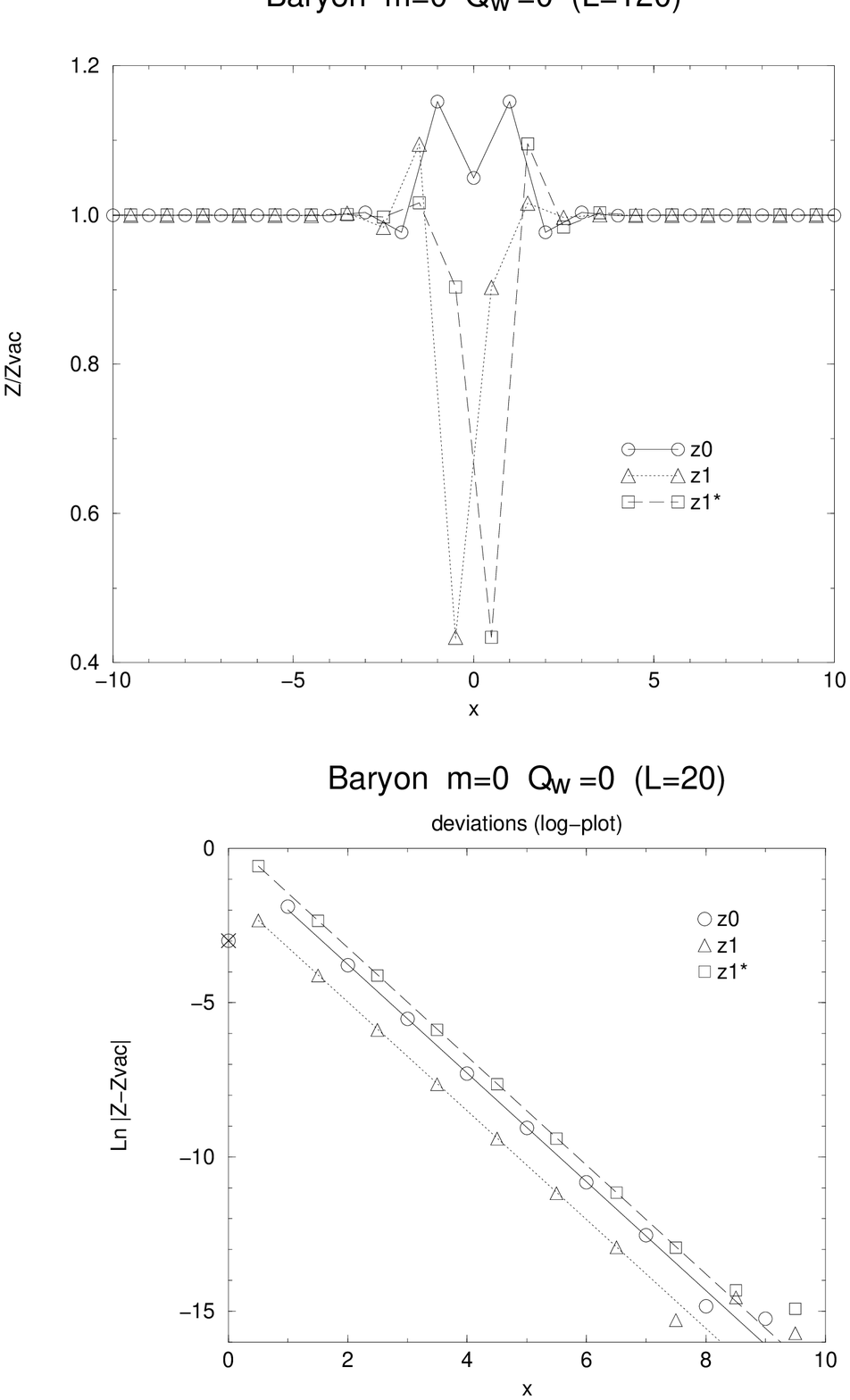}}}
  \lbfig{bar120-Q0m000}
\end{picture}
\caption{Numerical baryon configuration in the chiral limit.  All
  fields are real.  Top: the fields converge exponentially fast to
  $z_{\rm vac}$.  Bottom: we compare the numerical data (circles,
  triangles, squares) with the theory of Section \ref{baryon-chiral}
  (3 lines, cross for $z_0(0)$).}
\end{center}
\end{figure}
\newpage\thispagestyle{empty}
\begin{figure}[htb]
\begin{center}
  \setlength{\unitlength}{1cm}
\begin{picture}(10,21)
  \put(-2.7,0.0)
  {\mbox{\epsfysize=23.0cm\epsffile{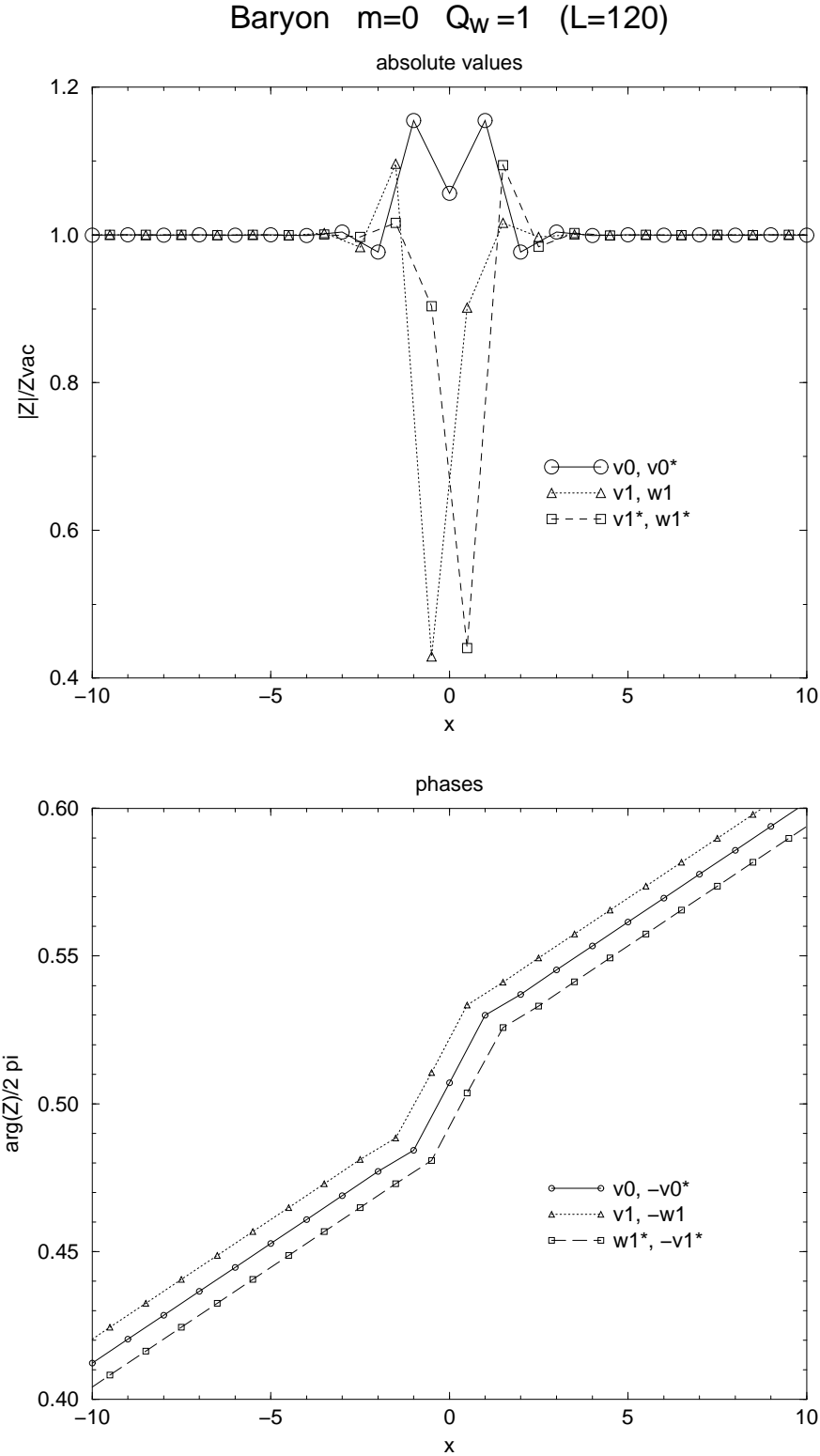}}}
  \lbfig{bar120-Q1m000}
\end{picture}
\caption{Numerical baryon in the chiral limit with winding number 
  $Q_w = 1$ (we only plot the vicinity of the baryon worldline).  The
  absolute values are very similar to the case $Q_w = 0$, but now the
  phases vary linearly away from the worldline.  The slope $\alpha$
  and the phase jump $\beta$ at the baryon are in good agreement with
  the theory of Section \ref{topobaryonchiral}: we find $2\pi / \alpha
  = 123.42$ and $\arg v_0(1) - \arg v_0(-1) = 0.046\approx 5.24
  \alpha$.}
\end{center}
\end{figure}
\newpage\thispagestyle{empty}
\begin{figure}[htb]
\begin{center}
  \setlength{\unitlength}{1cm}
\begin{picture}(10,21)
  \put(-2.7,0.0)
  {\mbox{\epsfysize=17.0cm\epsffile{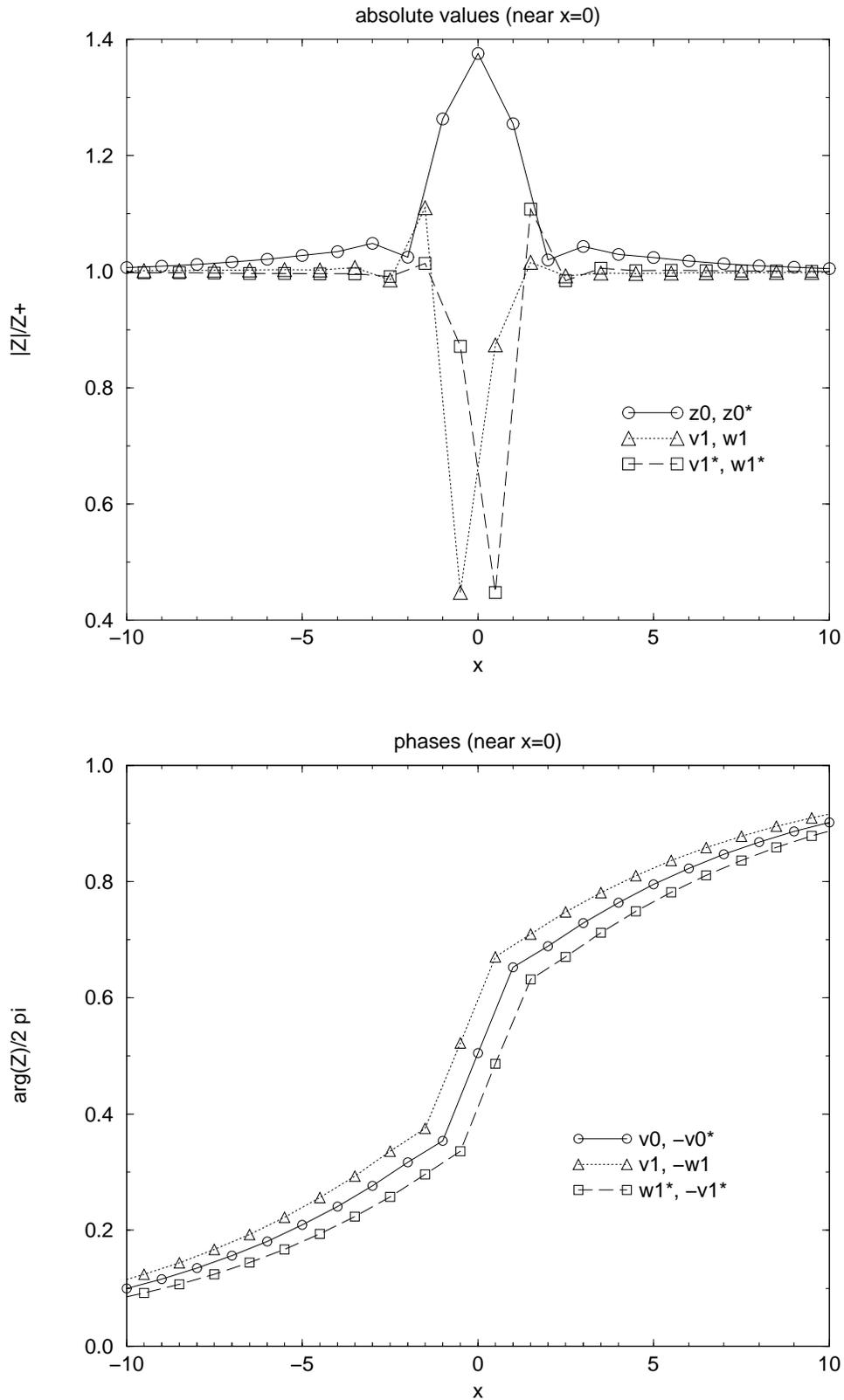}}}
  \lbfig{bar120-Q1m005}
\end{picture}
\caption{Numerical baryon with winding number $Q_w = 1$ and broken 
  chiral symmetry.  Away from the worldline, the fields converge
  exponentially fast to $z_+$.  The logarithms of the phases are
  linear in the range $3 < x < 55$, with slopes $0.1523 \leq \gamma
  \leq 0.1533$ in agreement with the theoretical value from
  Eqs.~\eqref{datam005}.}
\end{center}
\end{figure}
\newpage\thispagestyle{empty}
\begin{figure}[htb]
\begin{center}
  \setlength{\unitlength}{1cm}
\begin{picture}(10,21)
  \put(-2.7,-0.0) {\mbox{\epsfysize=17.0cm\epsffile{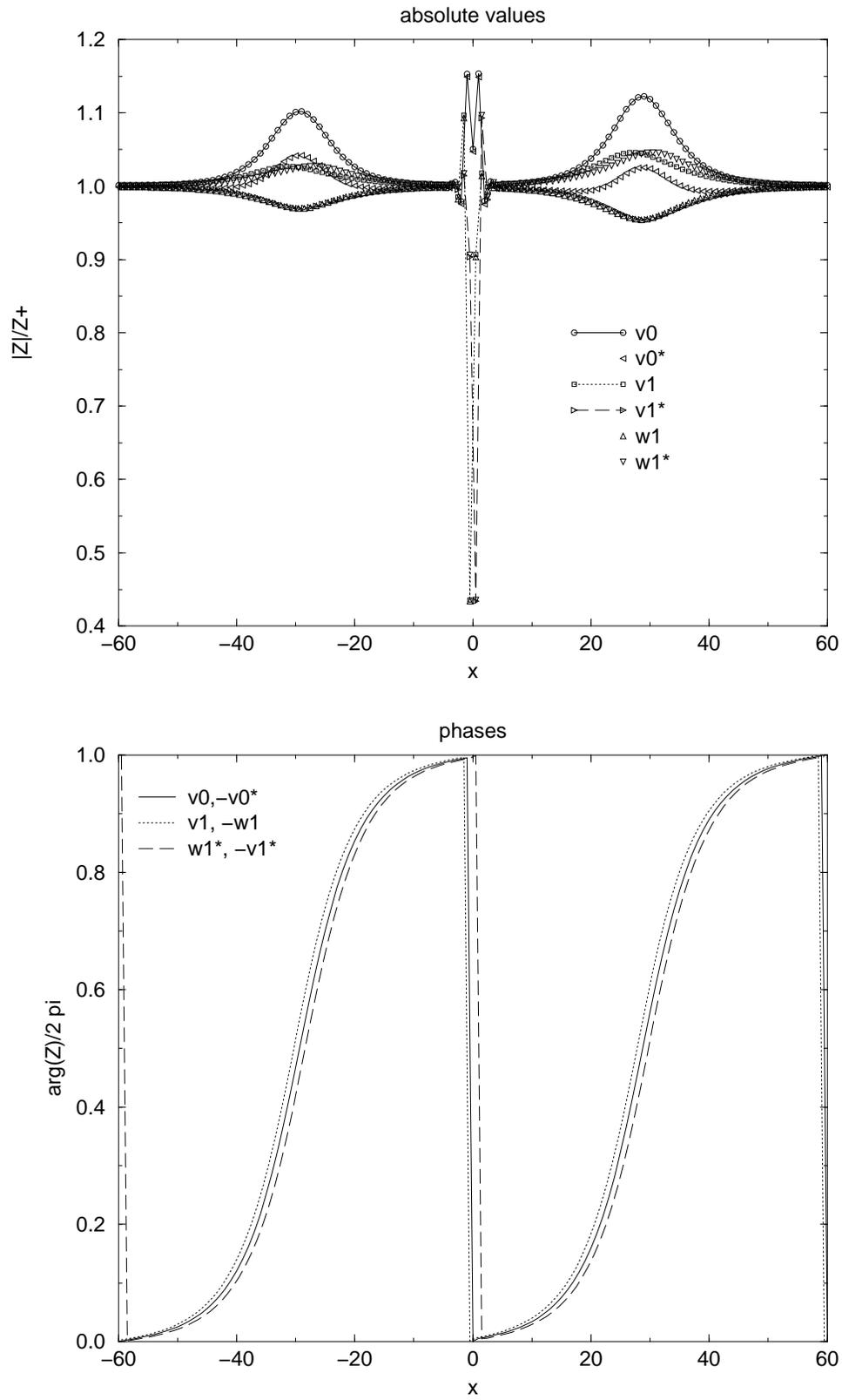}}}
  \lbfig{bar120-Q2m005}
\end{picture}
\caption{Numerical baryon with winding number $Q_w = 2$ and broken
  chiral symmetry.  The symmetry with respect to $x = 0$ is only
  approximate, due to a numerical loss of accuracy.}
\end{center}
\end{figure}
\newpage\thispagestyle{empty}
\begin{figure}[htb]
\begin{center}
  \setlength{\unitlength}{1cm}
\begin{picture}(10,12.5)
  \put(-2,-0.0) {\mbox{\epsfysize=12cm\epsffile{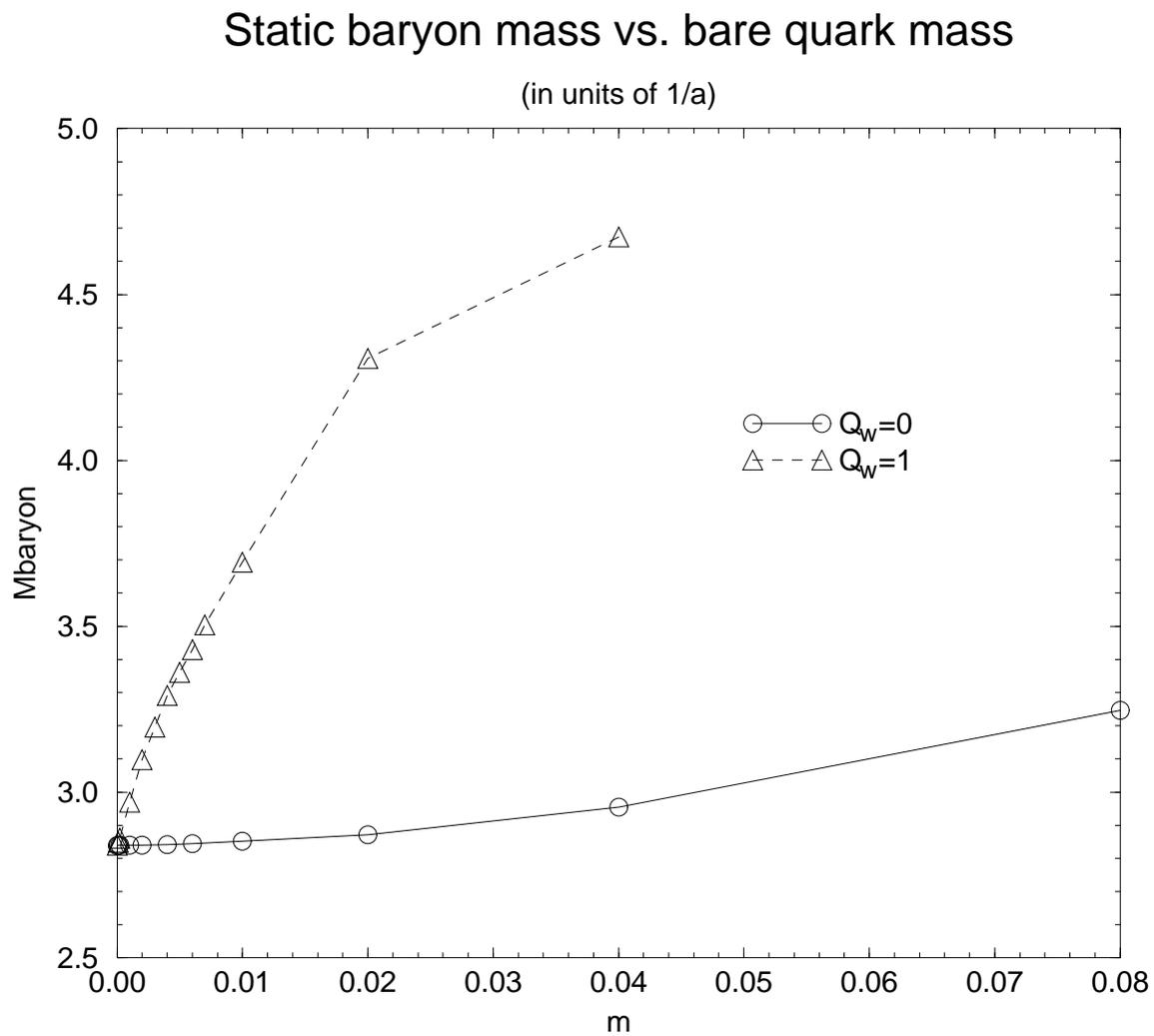}}}
  \lbfig{masses}
\end{picture}
\caption{Masses of static baryons as functions of the quark mass
  $am$, for the topological sectors $Q_w = 0$ and $Q_w = 1$.}
\end{center}
\end{figure}
\newpage\thispagestyle{empty}
\begin{figure}[t]
\begin{center}
  \setlength{\unitlength}{1cm}
\begin{picture}(10,12.5)
  \put(-2,-0.0) {\mbox{\epsfysize=12cm\epsffile{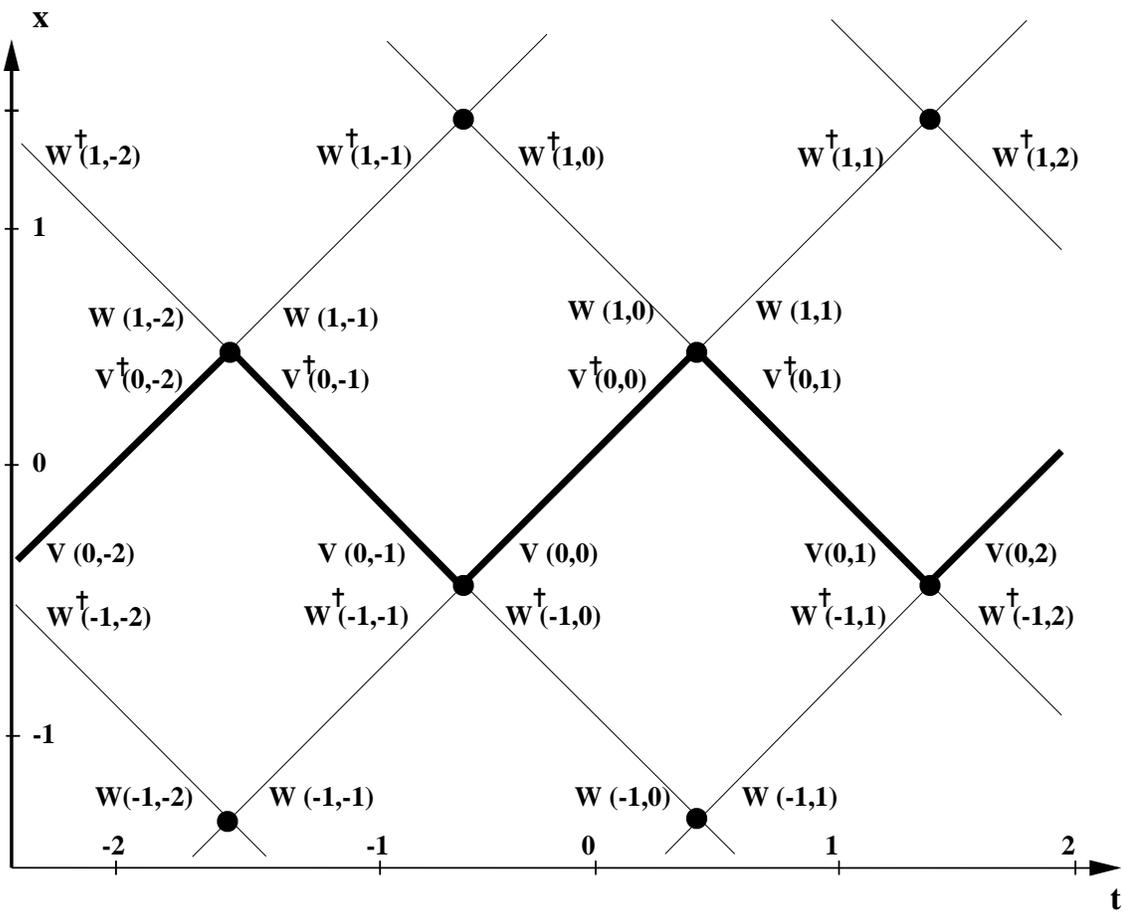}}}
  \lbfig{baryon-zigzag}
\end{picture}
\caption{Zigzag baryon string on a square lattice.}
\end{center}
\end{figure}
\newpage\thispagestyle{empty}
\begin{figure}[htb]
\begin{center}
  \setlength{\unitlength}{1cm}
\begin{picture}(10,21)
  \put(-3.0,0.5) {\mbox{\epsfysize=16cm\epsffile{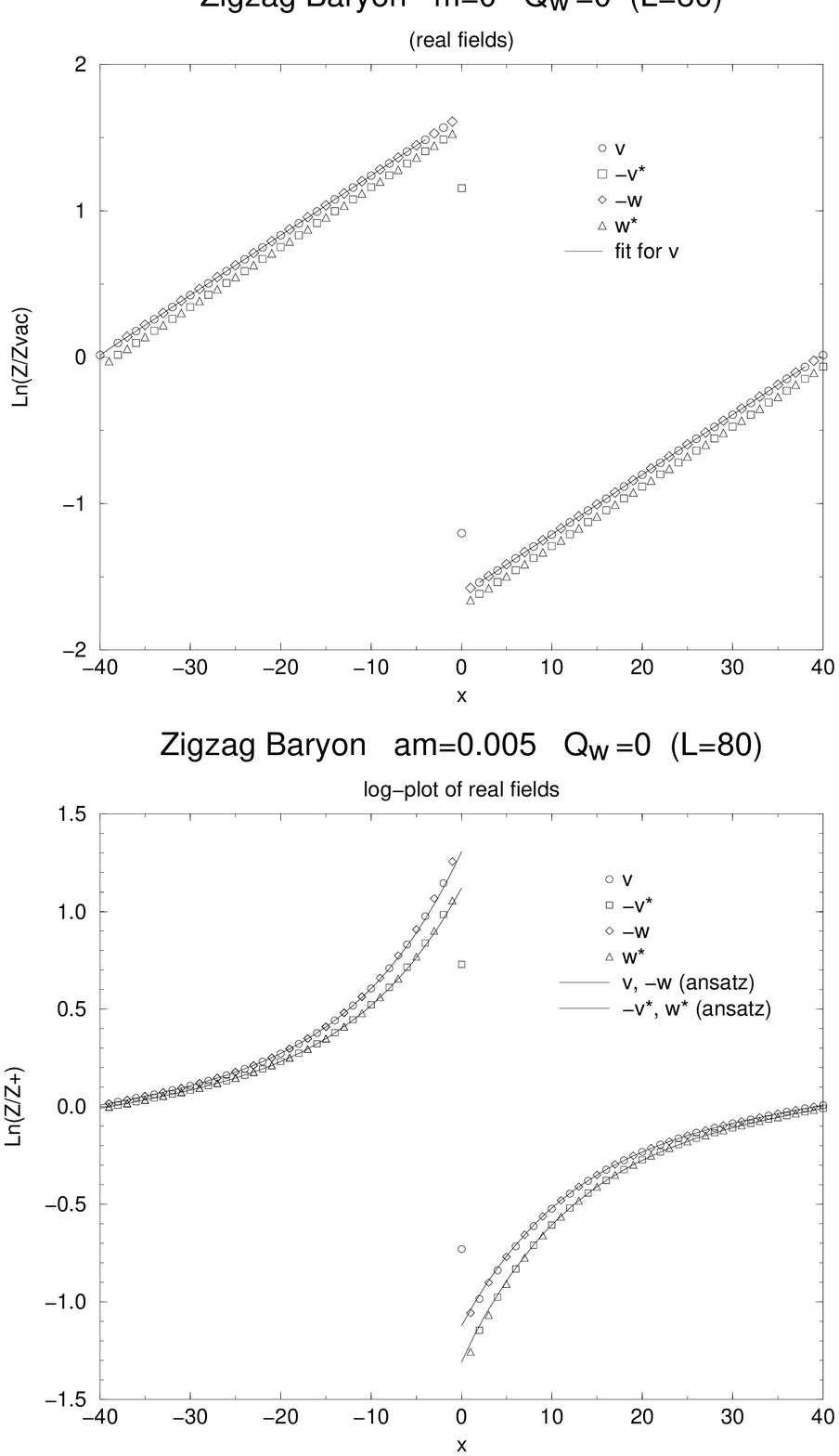}}}
  \lbfig{zig40-Q0}
\end{picture}
\caption{Numerical zigzag baryon for the case with chiral symmetry (top) 
  and without (bottom).  We plot the (real) fields on a logarithmic
  scale.  Top: a linear fit yields the slope $c_+ = 0.04$ and the
  field on the baryon $v(0) = 0.1735$, in excellent agreement with
  formulas (\ref{implicitc+},\ref{v(0)}).  Bottom: we use the
  exponential ansatz of Section \ref{zigbroken}, with coefficients
  $\eps_{\pm}$ fitted over the domain $|x'| < 30$.  The values $\eps_+
  = 0.062$, $\eps_- = -0.053$ are in good agreement with the
  analytical theory.}
\end{center}
\end{figure}
\newpage\thispagestyle{empty}
\begin{figure}[htb]
\begin{center}
  \setlength{\unitlength}{1cm}
\begin{picture}(10,21)
\put(-2.7,0.5)
{\mbox{\epsfysize=16.0cm\epsffile{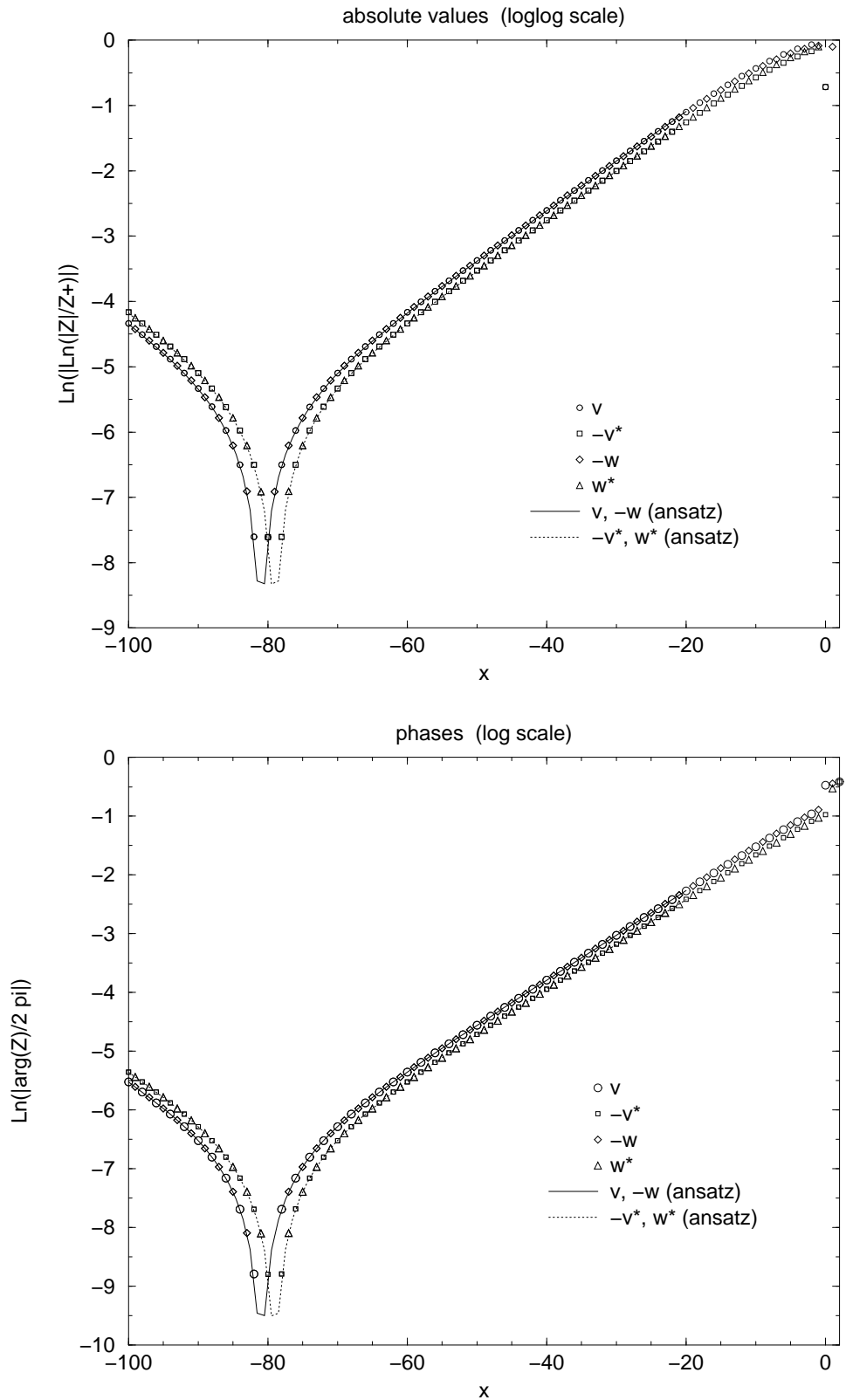}}}
\lbfig{zig80log}
\end{picture}
\caption{Numerical zigzag baryon with winding number $Q_w = 1$ and broken
  chiral symmetry.  We plot the real (top) and imaginary (bottom)
  parts of $\log(z/ z_+)$ on a logarithmic scale, together with a fit
  by the ansatz \eqref{ansatzzig} in the region $|x'| < 60$.  The
  coefficients take the values $\eps_+ \approx (3.48 + \mi 1.07) \times
  10^{-3}$, $\eps_- \approx -(3.00 + \mi 0.92) \times 10^{-3}$, and
  satisfy quite well the relation \eqref{mirror2}.}
\end{center}
\end{figure}
\end{document}